\documentclass[twocolumn]{aastex631}

\usepackage{amsmath}

\begin{document}

\title{Spatially-resolved TRGB JWST color-magnitude as a tool to measure fossil stellar metallicity gradients in disk galaxies: NGC 628}

\correspondingauthor{Avinash Ck}
\email{avinash@inaoe.mx}

\author[0000-0002-7507-7227]{Avinash Ck}
\affiliation{Instituto Nacional de Astrofísica, Óptica y Electrónica, Luis Enrique Erro 1, Tonantzintla 72840, Puebla, Mexico}

\author[0000-0002-4677-0516]{Divakara Mayya}
\affiliation{Instituto Nacional de Astrofísica, Óptica y Electrónica, Luis Enrique Erro 1, Tonantzintla 72840, Puebla, Mexico}

\author[0000-0002-7922-8440]{Alessandro  Bressan}
\affiliation{Scuola Internazionale Superiore di Studi Avanzati, Via Bonomea, 265, I-34136, Trieste, Italy}

\author[0000-0002-8351-8854]{Andres Jairo Alzate Trujillo}
\affiliation{Centro de Estudios de Física del Cosmos de Aragón (CEFCA), Plaza San Juan, Teruel, Spain}

\begin{abstract}
We use archival JWST/NIRCam images in the F115W, F150W, and F200W filters to measure the Tip of the Red Giant Branch (TRGB) magnitudes across the disk of the late-type spiral galaxy NGC\,628. In this exploratory study, we demonstrate how the metallicity-dependence of TRGB magnitudes in the near-infrared (NIR) filters can be exploited by making use of the theoretical isochrones to determine metallicities of the fossil 10~Gyr old population over kiloparsec scales without being affected by the age-metallicity-reddening degeneracy. We obtain a smooth metallicity gradient that decreases from $Z$=0.003 in the central regions to $Z$=0.002 in the external parts, with a typical error on $Z$  of 0.0002. The extinction variation across the disk caused by the diffuse interstellar dust is spiky with a median value of $A_V$=0.12~mag. We propose that the large bubbles in the disks of galaxies offer dust-free lines of sight for measuring the TRGB magnitudes, and hence the distance to galaxies, to an accuracy that is as good as that of the halo populations. Using the Phantom Void, we obtain a TRGB distance modulus of 29.81$\pm0.05(\rm stat)\pm0.06(\rm sys)$~mag for NGC\,628, which agrees well with the most recent PNLF measurement of 29.89$^{+0.06}_{-0.09}$ for this galaxy.
\end{abstract}

. 
\section{Introduction}

\begin{figure*}
    \centering
\includegraphics[width=0.5\textwidth]{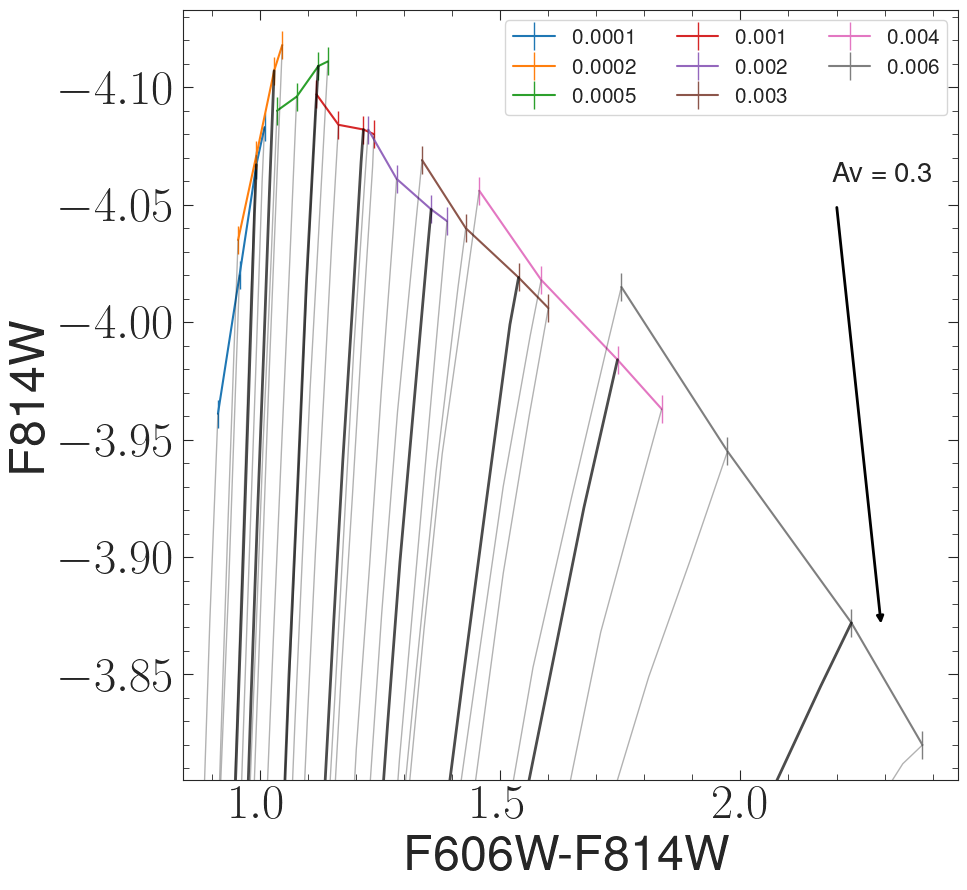}%
\includegraphics[width=0.5\textwidth]{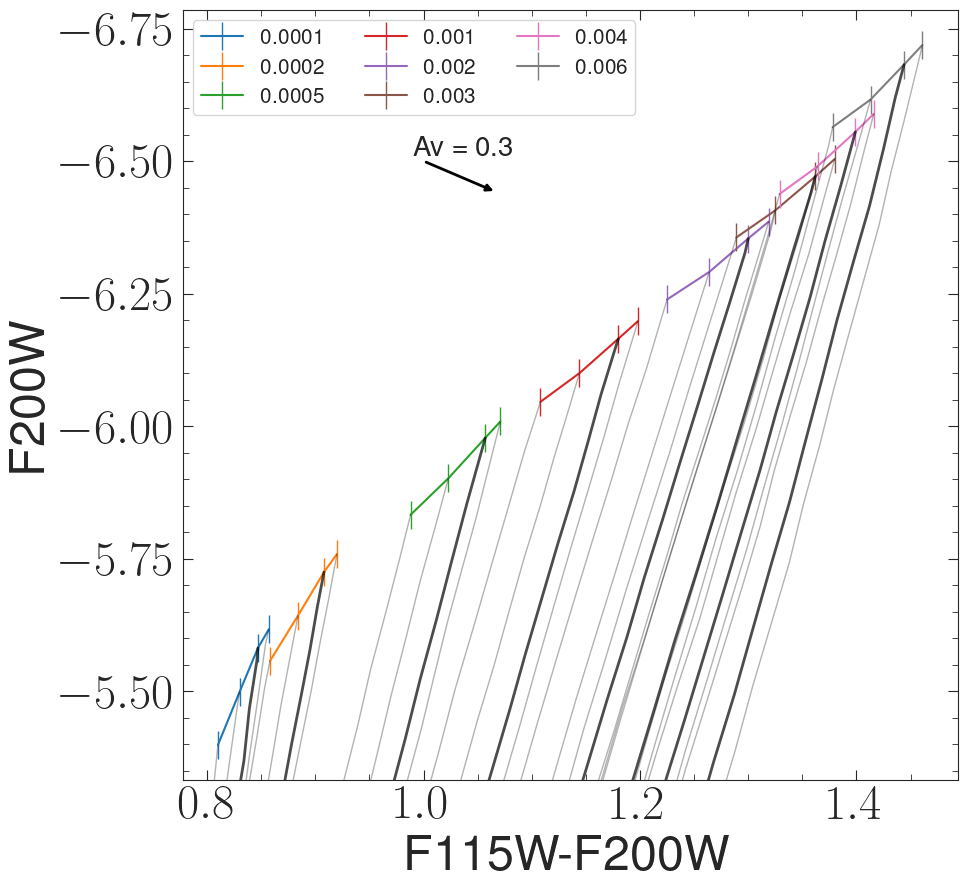}
    \caption{Sensitivity of the TRGB magnitude and colors to age and metallicity in the HST F814W (left) and JWST F200W (right) filters. The plots show PARSEC 1.2S isochrone (almost vertical lines) zoomed around the TRGB stage for 8 metallicities (see the inset box) and 4 ages (4, 6.3, 10, and 13 Gyr). A thick black line shows the 10 Gyr isochrone at each metallicity, which allows us to visualize the sensitivity of the plotted quantities on metallicity. The age dependence of the TRGB magnitude can be inferred from the length of the line joining the same metallicity points. The reddening vector is shown.}
    \label{fig:parsec_iso}
\end{figure*}

The tip of the red giant (TRGB) method is one of the best methods for measuring extragalactic distances \citep{Riess_2022, Riess_2024, freedman_2024}. TRGB is well-understood theoretically \citep[e.g.][]{trgb_theory_1, trgb_theory_2} and represents the old ($\sim$4--10~Gyr), metal-poor (Z=0.0001--0.006) populations of galaxies. The TRGB magnitude, however, depends on age, metallicity, and the extinction along the line of sight. Theoretical stellar evolution models \citep[e.g. PARSEC;][]{parsec} reveal that age and metallicity dependence is low at the optical wavelengths, with variations below $\sim$0.1 magnitude in the Hubble Space Telescope (HST) F814W filter for a metallicity range of Z=0.0001--0.004 (Figure~\ref{fig:parsec_iso} (left) and Table \ref{tab:trgb}), the typical range of metallicities of the halo populations. This has led to the use of TRGB magnitudes in the HST-F814W filters as standard candles for determining the Hubble constant at high precision (\citep{Riess_2022}. On the other hand, at longer wavelengths, now accessible with the Near InfraRed Camera (NIRCam) onboard the James Webb Space Telescope (JWST), the TRGB magnitude depends both on metallicity and age, with the metallicity dependence stronger of the two (Figure~\ref{fig:parsec_iso}, right, and Table \ref{tab:trgb}). Use of F090W magnitude \citep{trgb_4258, trgb_fornax, trgb_virgo}, and correcting the longer wavelength magnitudes for the metallicity dependence, based on color-dependent empirical calibrations \citep{trgb_jwst_theory,newman_2024}, are two of the most commonly used strategies to make use of the increased sensitivity and spatial resolution of JWST as compared to the HST.

 One noteworthy feature of the NIR Color Magnitude Diagram (CMD) shown in Figure~\ref{fig:parsec_iso} (right) is the near orthogonal dependence of reddening vector to the dependence on age and metallicity. This near orthogonality, together with the relatively low sensitivity of TRGB magnitudes to age, allows the determination of metallicity and reddening without the problems of degeneracy in these two parameters.  Thus, leveraging this dependence of TRGB magnitudes in longer wavelength NIRCam filters based on the theoretical isochrones, rather than empirical calibrations, could provide a novel method to obtain the interstellar reddening and metallicity of the old stellar populations in galaxies. Successful testing of the method opens up the possibility of mapping the extinction and metallicity of the old disks of galaxies, right from the time of the disk formation.
The availability of extensive data on the disks of spiral galaxies from FEAST \citep{feast} and PHANGS \citep{phangs} in the first 2 cycles of JWST observations makes exploring this dependence on the disk populations even more compelling. In this work, we carry out such a study, taking NGC628 as an illustrative case.

NGC628 is a late-type spiral galaxy \citep{late_type}, currently forming stars at 1.7 $M_{\odot}yr^{-1}$ \citep{muse_sfr}. It is included in two recent JWST samples: PHANGS and FEAST, mainly because it represents a star-forming disk at nearly face-on orientation. Matching the observed TRGB magnitudes with that from the theoretical isochrones, which requires a precise Distance Modulus (DM), is fundamental for exploring the NIR CMDs. NGC\,628 has several distance measurements, including from the TRGBs from HST/F814W filter data \citep{trgb_hst_1, trgb_hst_2, trgb_hst_3}, giving DM ranging from 29.68 to 30.04. In a recent study, \cite{pnlf1} obtained a new DM=29.89$^{+0.06}_{-0.09}$ using Planetary Nebulae Luminosity Function (PNLF) from the MUSE dataset. We re-evaluate these distances using the TRGBs in the JWST F115W and F200W images.

\begin{deluxetable}{lcccc}
\tabletypesize{\scriptsize}
\tablecaption{Distance Modulus (DM) variation with age, metallicity, and extinction}
\label{tab:trgb}
\tablehead{\colhead{Parameters} & \colhead{F814W} & \colhead{F115W} & \colhead{F150W} 
& \colhead{F200W}}
\colnumbers
\startdata
$\left.\Delta(DM)_{Age}\right|_{Z=0.0001}$ & 0.122 & 0.171 &  0.203 & 0.218\\
$\left.\Delta(DM)_{Age}\right|_{\textbf{Z=0.003}}$  & \textbf{0.063} & \textbf{0.057} &  \textbf{0.1}   & \textbf{0.148}\\
$\left.\Delta(DM)_{Age}\right|_{Z=0.006}$  & 0.195 & 0.072 &  0.106 & 0.155\\
$\left.\Delta(DM)_{Z}\right|_{Age=4Gyr}$   & 0.136 & 0.597 &  0.878 & 1.165\\
$\left.\Delta(DM)_{Z}\right|_{Age=6.3Gyr}$ & 0.151 & 0.534 &  0.815 & 1.117\\
$\left.\Delta(DM)_{Z}\right|_{\textbf{Age=10Gyr}}$   & \textbf{0.237} & \textbf{0.503} & \textbf{0.786} & \textbf{1.1}\\
$\left.\Delta(DM)_{Z}\right|_{Age=13Gyr}$  & 0.298 & 0.498 & 0.781 & 1.102\\
$\Delta(DM)_{Av}$ & 0.299 & 0.21&0.144 &0.098\\
\enddata
\tablecomments{$\Delta(DM)_{age}$ is the difference between the maximum and minimum DMs due to age variations between 4 and 13 Gyr, at a fixed (indicated) metallicity. $\Delta(DM)_Z$ is the difference between the maximum and minimum DMs due to metallicity variations between Z=0.0001 and 0.006, at a fixed (indicated) age. $\Delta(DM)_{A_V}$ is the change in DM due to a foreground extinction amounting to $A_V=0.5$ mag.}
\end{deluxetable}

In Section~\ref{sec:data}, we describe the data utilized in this work and the stellar photometry routine. In Sections~\ref{sec:cmd} and \ref{sec:method}, we describe the filters used for TRGB estimation and the methodology. The results from this study are discussed in Section~\ref{sec:result}, where we also discuss the TRGB distance obtained in this study with the previous determinations, and Section~\ref{sec:conclusions} gives a conclusion of this study.

\begin{deluxetable*}{lcccc}
\tabletypesize{\scriptsize}
\tablecaption{Description of imaging data utilized in the work}
\tablewidth{0pt}
\colnumbers
\tablehead{\colhead{NIRCam Filter}&\colhead{Exposure ids}&\colhead{Date}&\colhead{Header Exposure time}&\colhead{Corrected Exposure time}}%
\startdata 
F115W & jw01783004003\_02101\_0000[1-4]\_nrc$\dagger$ &2022-07-29& 365s$\times$4 & 322.1031s$\times$4 \\
F115W & jw01783004004\_02101\_0000[1-4]\_nrc$\dagger$ &2022-07-29& 365s$\times$4 & 322.1031s$\times$4 \\
F150W & jw01783004005\_02101\_0000[1-4]\_nrc$\dagger$ &2022-07-29& 118s$\times$4 & 107.3677s$\times$4 \\
F150W & jw01783004006\_02101\_0000[1-4]\_nrc$\dagger$ &2022-07-29& 118s$\times$4 & 107.3677s$\times$4 \\
F200W & jw01783004007\_02101\_0000[1-4]\_nrc$\dagger$ &2022-07-29& 365s$\times$4 & 322.1031s$\times$4 \\
F200W & jw01783004008\_02101\_0000[1-4]\_nrc$\dagger$ &2022-07-29& 365s$\times$4 & 322.1031s$\times$4 \\
\enddata
\tablecomments{$\dagger$ All 8 detector nrc[a1-a4] and nrc[b1-b4] were used.}
\label{tab:data}
\end{deluxetable*}

\section{Data and stellar photometry}
\label{sec:data}
\subsection{Dataset}
In this study, we analyzed publicly available near-infrared short-wavelength (NIRCam SW) and mid-infrared (MIRI) data of NGC\,628 from the JWST-FEAST\footnote{JWST probes Feedback in Emerging extrAgalactic Star clusTers} \citep[Proposal ID 1783,][]{feast} observations. Specifically, we used the NIRCam filters F115W, F150W, F200W, and the MIRI filter F770W. The NIRCam and MIRI images have spatial samplings of 0.031"/pixel and 0.11"/pixel, respectively. We obtained raw NIRCam exposures (\textit{\_uncal} files) through the MAST Portal and reprocessed them with the JWST pipeline \citep{jwstpipeline} using reference files from the \textit{jwst\_1293.pmap} context obtained from CRDS\footnote{\url{https://jwst-crds.stsci.edu/}}. The description of the NIRCam data is provided in Table \ref{tab:data}. The data were obtained using a FULLBOX$\times$4 TIGHT dither pattern, utilizing all 8 NIRCam SW detectors and generating 8 exposures per detector, covering a field of view of approximately $6'\times2.3'$ centered on the galaxy.
\subsection{Resolved stellar photometry}
\begin{figure}
    \centering
\includegraphics[width=\columnwidth]{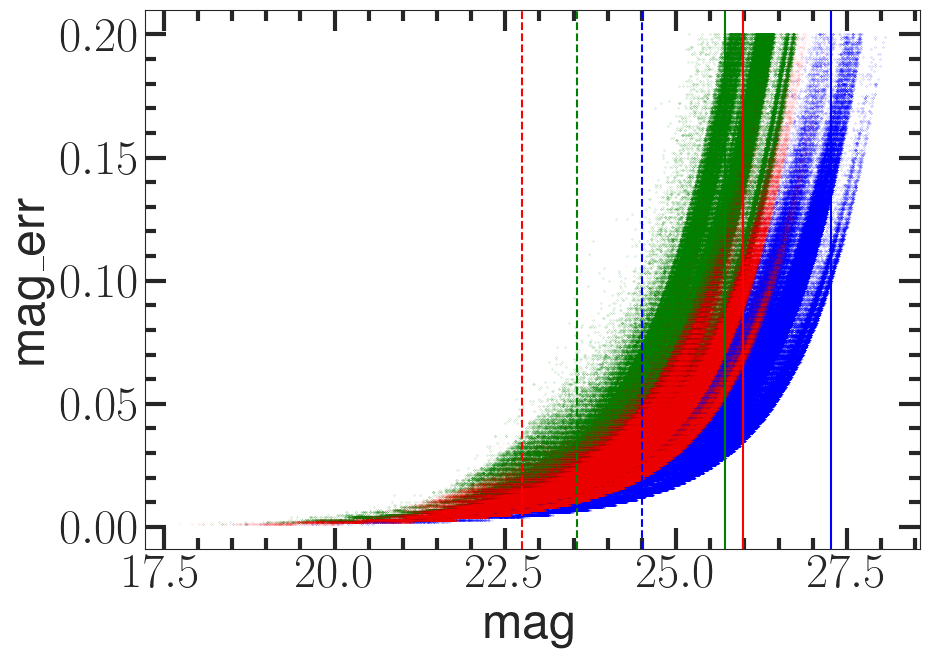}
    \caption{Magnitude error plotted against DOLPHOT magnitude in three filters over the entire FoV of NGC 628. Blue, green, and red points indicate the photometry of individual stars in F115W, F150W, and F200W filters, respectively. The completeness magnitudes $m_{50}$ for the deepest and shallowest zones in the galaxy are shown by solid and dashed vertical lines, respectively, with the colors of the lines blue, green, and red corresponding to F115W, F150W, and F200W.}
    \label{fig:mag_err}
\end{figure}
For photometry, we utilized the DOLPHOT software's NIRCam\footnote{\url{http://americano.dolphinsim.com/dolphot/nircam.html}} module \citep{dolphot_1, dolphot_2}. DOLPHOT is an iteratively subtracted PSF photometry tool optimized for crowd-field photometry. We employed the DOLPHOT routine for PSF photometry as per the instructions mentioned in the DOLPHOT NIRCam Manual\footnote{\url{http://americano.dolphinsim.com/dolphot/dolphotNIRCam.pdf}}. We adopted the DOLPHOT parameters optimized with the Early Release Science (ERS) program \citep{ers,Weisz_2024}. The Stage 3 F200W \textsf{(\_I2D)} drizzled image was used as the reference image for the photometry. The photometry was performed on the \textsf{CRF} files across filters (F115W, F150W, F200W), simultaneously, on a detector-by-detector basis, optimized for memory usage. We used the photometry of stars from overlapping regions from different detector exposure sets originating from dithering to check the photometric quality. For generating a clean stellar photometric catalog we performed quality cuts on the DOLPHOT output photometry catalog as per \cite{Warfield_2023}. Specifically, we selected sources with \textit{Object\_Type $\leq2$, Crowding $\leq0.5$, Sharpness$^2\leq0.01$, Flag $\leq 2$ and SNR $\geq5$}, and obtained a sample of $\sim$1,100,000 stars in the entire field of view.

The results of stellar photometry in F115W, F150W, and F200W filters are shown in blue, green, and red colors, respectively in Figure~\ref{fig:mag_err} over the entire FoV. The errors are as defined in DOLPHOT, which include the detector noise, Poisson noise due to the source and background, and the error due to the PSF fitting process. As expected, the error increases at fainter magnitudes. For a given photometric error, the F115W image detects fainter stars than the other two filters. The relatively lower exposure time of the F150W image makes the images in this filter least sensitive to faint magnitudes among the three filters. The limiting magnitude varies over the FoV due to varying background brightness and crowding in different parts of the image, which is the reason for the spread and multiple stripes in each filter. The 50\% completeness limit $m_{50}$ is shown (vertical lines) for two extreme crowding values in each of the three filters. We explain below the procedure we have followed to obtain $m_{50}$ in each filter at different zones of the image.

\begin{figure}
    \centering
\includegraphics[width=\columnwidth]{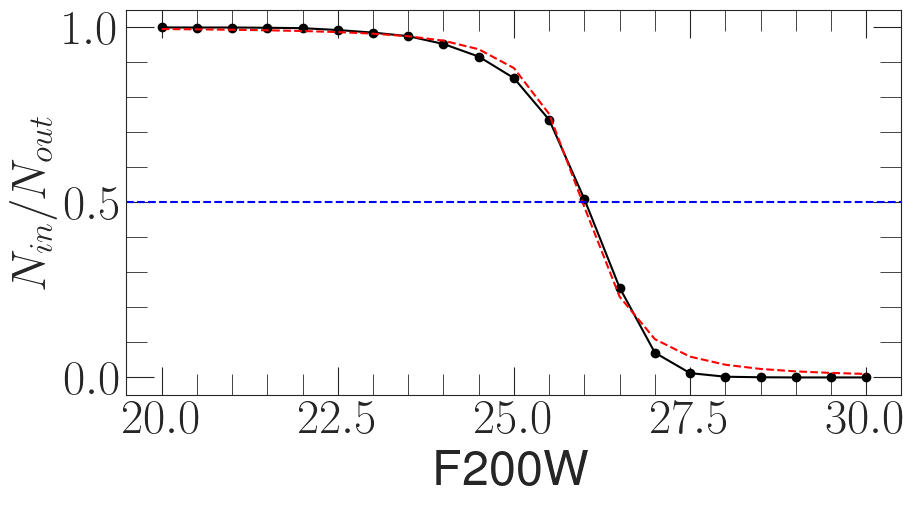} 
\includegraphics[width=\columnwidth]{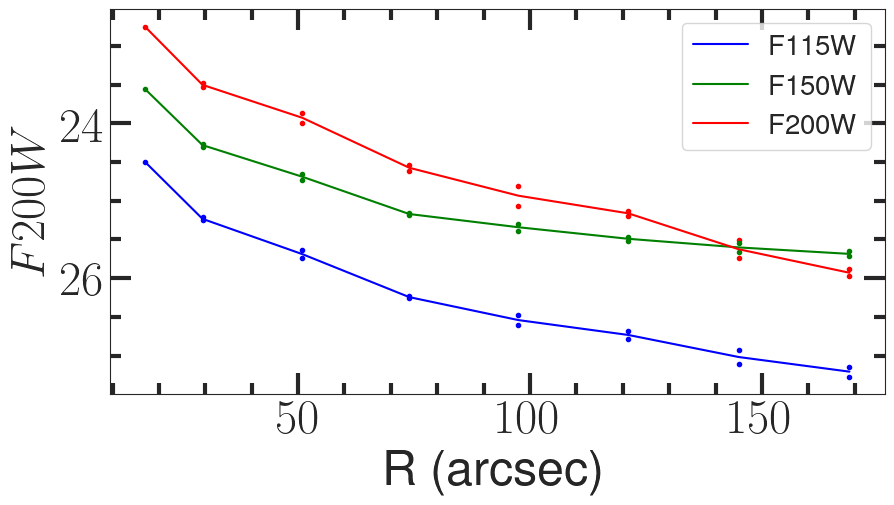} 
    \caption{(Top) Pritchet function fits a completeness curve for F200W (black curve with points every 0.5~mag) for an illustrative region in the external disk of the galaxy. The intersection of the blue horizontal line with the Pritchet function indicates the 50\% completeness magnitude. 
     (Bottom) Radial dependence of $m_{50}$ in filters F115W (blue), F150W (green), and F200W (red).}
    \label{fig:m50}
\end{figure}

\begin{figure*}
    \centering
\includegraphics[width=\textwidth]{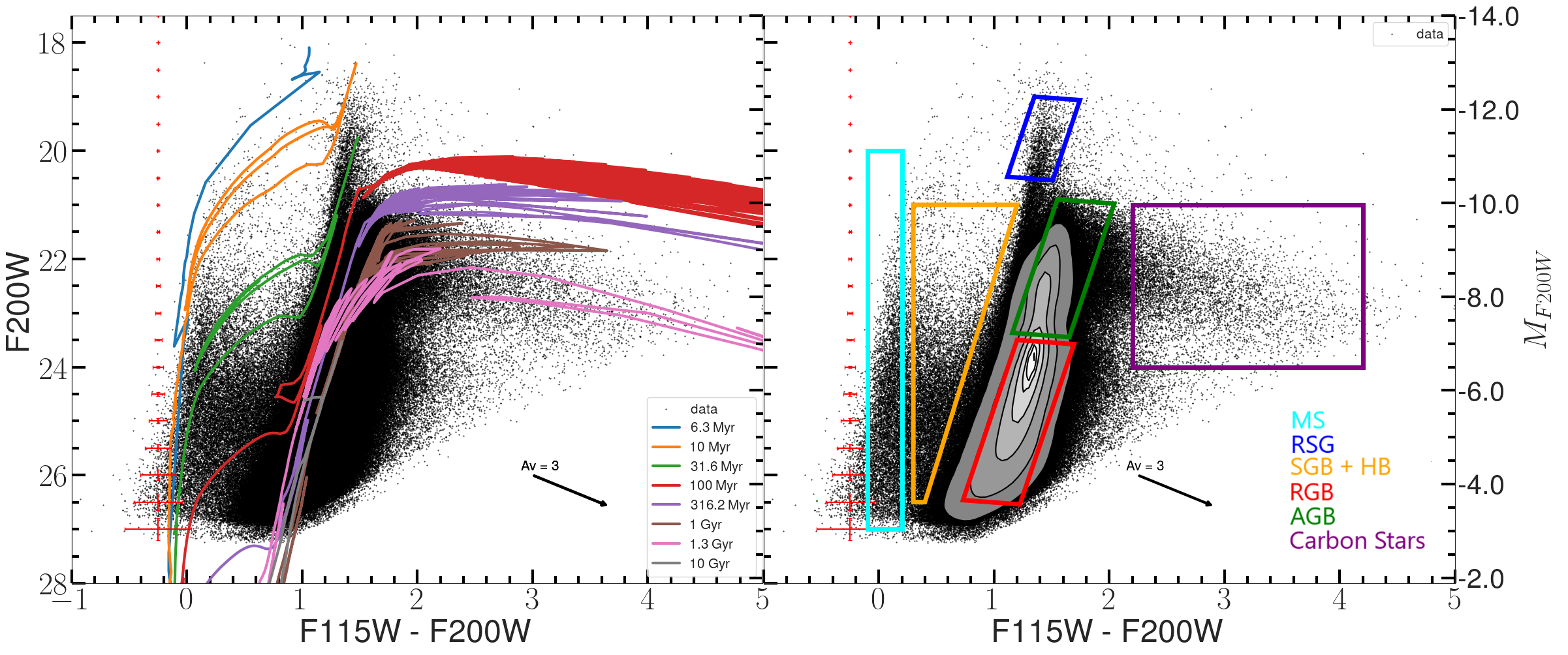}
    \caption{NIRCam F115W$-$F200W CMD over the entire galaxy disk of NGC\,628. In the left panel, we superpose the PARSEC isochrones of selected ages, with the younger (log(age/yr)$\leq$9) isochrones being metal-rich (Z=0.02) than the older ones (Z=0.002). In the right panel, we mark the most prominent and easily identifiable stages of stellar evolution in the CMD of a typical disk and show the Kernel densities with contours at 10\%, 25\%, 50\%, 75\%, 90\%, 95\%, and 99\%. The reddening vector is shown by the arrow.}
    \label{fig:cmd}
\end{figure*}

\subsection{Completeness Analysis}

Over the relatively large FoV of the JWST images, the surface brightness and crowding due to the over-density of stars vary significantly, resulting in variations in the limiting magnitude of detection. To illustrate this, we performed completeness tests using DOLPHOT in different zones (see grid regions in Section~4.3) representing the entire range of stellar densities and crowding in this galaxy. We populated 10,000 stars of a fixed magnitude onto the JWST image in a given filter in each zone and then retrieved them using the same quality cuts and filtering process as the observed data. The magnitudes are varied from 20 to 30 at intervals of 0.5 magnitude. For obtaining the 50\% completeness magnitude ($m_{50}$), we used the Pritchet function \citep{pritchet_1, pritchet_2, pritchet_3, pritchet_4} given by
\begin{equation*}
f(m) = \dfrac{1}{2}\left[1-\dfrac{\alpha(m-m_{50})}{\sqrt{1+\alpha^2(m-m_{50})^2}}\right]
\end{equation*} 
to fit the completeness curves, where $\alpha$ is a fitting constant. A completeness curve is essentially the ratio of the number of stars retrieved and the number of stars simulated ($f(m) = N_{out}/N_{in}$)  as a function of the magnitude of the simulated stars, which is shown in Figure~\ref{fig:m50} (top) for the F200W image for a region in the outer part of the galaxy over a square area of 24~arcsec a-side. A $m_{50}$=26.0~mag is derived from this plot.

In Figure~\ref{fig:mag_err}, the deepest $m_{50}$ corresponds to the furthest region from the galaxy center with the least stellar density, while the most shallow $m_{50}$ corresponds to the grid containing the galaxy center. In this zone, the stellar density is too high to be able to resolve the entire population, giving rise to a higher crowding as well as background from the unresolved populations. In regions with lower stellar densities, $m_{50}$ is determined by the detection limit, making F150W have the lowest $m_{50}$ due to the lowest exposure time. However, crowding dictates $m_{50}$ in regions with higher stellar densities. 
In Figure~\ref{fig:m50} (bottom), we show the completeness magnitude $m_{50}$ as a function of galactocentric radius in F115W, F150W and F200W filters. As expected, $m_{50}$ becomes brighter towards the center in all filters.
Stars below the detection limit (e.g. low-mass main sequence stars) are in general red objects that give a higher background in F200W as compared to the other two filters, which is the reason for systematically lowering $m_{50}$ in this filter as compared to the other two filters.

\section{CMD and isochrones}
\label{sec:cmd}

Figure~\ref{fig:cmd} (left) shows F115W$-$F200W vs F200W color-magnitude diagrams (CMD) generated by combining photometry from all exposures across detectors and cross-matching between detectors with a crossmatch radius of 0.06 arcsec (2 pixels in NIRCam Short Wavelength). We have overplotted PARSEC 1.2S \citep{parsec} isochrones at selected ages. The NIRCam magnitudes are in the Vega system, and the isochrones are adjusted for a foreground extinction of $A_V=0.19$ \citep{foreground_av} and a distance modulus of 29.81~mag (Estimated in this work). The extinction ratios for F115W, F150W, and F200W were obtained from \cite{extinction_law}.

A CMD of the entire disk is far from a single metallicity and age population. Besides, internal extinction and its variation across the disk smear any sharp feature along the reddening vector. Nevertheless, many well-known evolutionary phases spanning a range of ages can be identified in this figure. To identify these different features, we mark broad zones that contain known stellar types in the right panel. We show contours of stellar densities in the most populated part of the CMD. The marked zones correspond to Main Sequence (MS), Red Super Giant (RSG), Subgiant Branch (SGB), Red Giant Branch (RGB), Asymptotic Giant Branch (AGB) stars, and Carbon-rich AGB stars in their thermally-pulsing (TP-AGB) phase. 
We observe that the majority of the stars belong to the RGB, and AGB phases, with less than 10\% of the stars in the MS, SGB,
RSG, and Carbon star phases. By superposing the isochrones corresponding to different metallicities 
we infer that Z=0.02 better represents the RSG stars, while Z=0.002 better represents AGB and Carbon stars. This is in agreement with the theoretical expectations, as RSG stars indicate the presence of younger stellar populations ($<50$~Myr) that are expected to be metal-rich as compared to the much older ($>$1~Gyr) AGB stars. The spread in the RGB colors is attributed to age, metallicity, and extinction variation in the galaxy, as we illustrate in the next section.
This spread completely prevents us from identifying the stars in the TRGB phase in the CMD for the entire galaxy. In the next section, we describe the methodology that we have followed to identify the TRGB stars in the F115W$-$F200W vs F200W CMD.

\section{Methodology \& Analysis}
\label{sec:method}

In this section, we describe a method for identifying the TRGB in the F115W$-$F200W vs F200W CMD of the disk populations and using them for the measurement of useful physical parameters of the disk.

From the JWST-FEAST NIRCam data, we use the F200W filter for TRGB magnitude measurement as it offers the best compromise between intrinsic TRGB magnitude and sensitivity among the F115W, F150W, and F200W filters. Additionally, we utilize the F115W-F200W color, as it provides the broadest wavelength baseline among the three filters, thus offering the maximum sensitivity to variations in metallicity and reddening. Figure~\ref{fig:parsec_iso} illustrates the age and metallicity dependence on TRGB magnitude in the F115W$-$F200W vs. F200W CMD using the PARSEC 1.2S isochrones. We can observe that in F200W, the TRGB magnitude varies by almost 1 magnitude for metallicity variation from 0.0001 to 0.006, whereas the variation is $\sim$0.2 for an age variation between 4--10~Gyrs. This large variation makes the F200W filter undesirable for empirically-based calibrations of the absolute magnitude, but excellent for using TRGB magnitudes to explore the metallicity and age of the disk populations with the use of theoretical isochrones. It can be noted from Figure~\ref{fig:parsec_iso} that the direction of variation of the TRGB in the CMD due to age and metallicity is different from that due to extinction in the NIR CMD. More specifically, in the HST-F814W filter, increasing metallicity and increasing $A_V$ both make the TRGB dimmer. However, in the JWST-F200W filter, increasing metallicity brightens the TRGB, while increasing $A_V$ still causes it to dim. This opposite behavior of metallicity and $A_V$ makes the JWST F115W$-$F200W vs. F200W CMD an excellent diagnostic tool to determine metallicity and $A_V$ of the disk regions using TRGBs as tracer populations.

\begin{figure*}
\centering
\includegraphics[width=\textwidth]{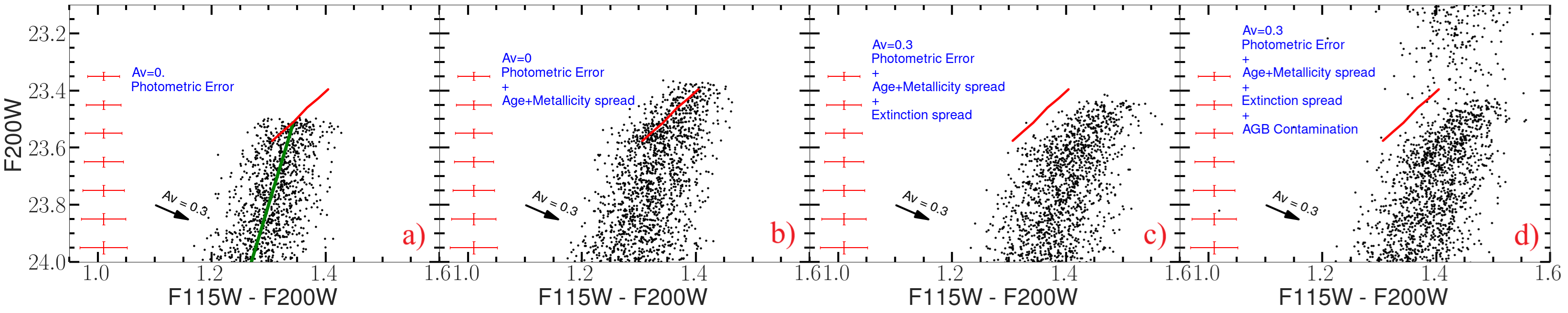}
\caption{Simulated CMDs using PARSEC models. (a) 10 Gyr, Z=0.002 population (green line) with simulated photometric error (points). The vertical sequence of red bars on the left indicates the typical photometric error as a function of magnitude. The thick red line is the locus of age-metallicity variation for $9.8\leq\log($Age$)\leq10$ and $0.002\leq$Z$\leq 0.003$, from bottom-left to top-right. (b) Same as (a) but with an age-metallicity spread corresponding to the thick red line. (c) Same as (b), but with a mean $A_V$=0.3 and a spread of 0.1~mag. (d) same as (c), but with an added AGB component, modeled using a Gaussian distribution.}

\label{fig:sim_CMD}
\end{figure*}

\begin{figure*}
\centering
\includegraphics[width=\textwidth]{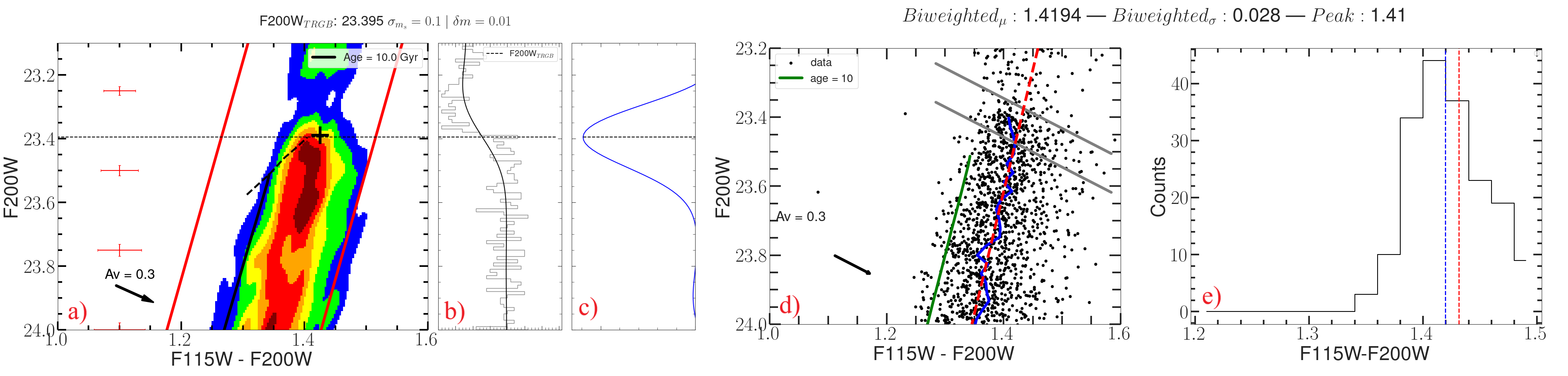}\\
\includegraphics[width=\textwidth]{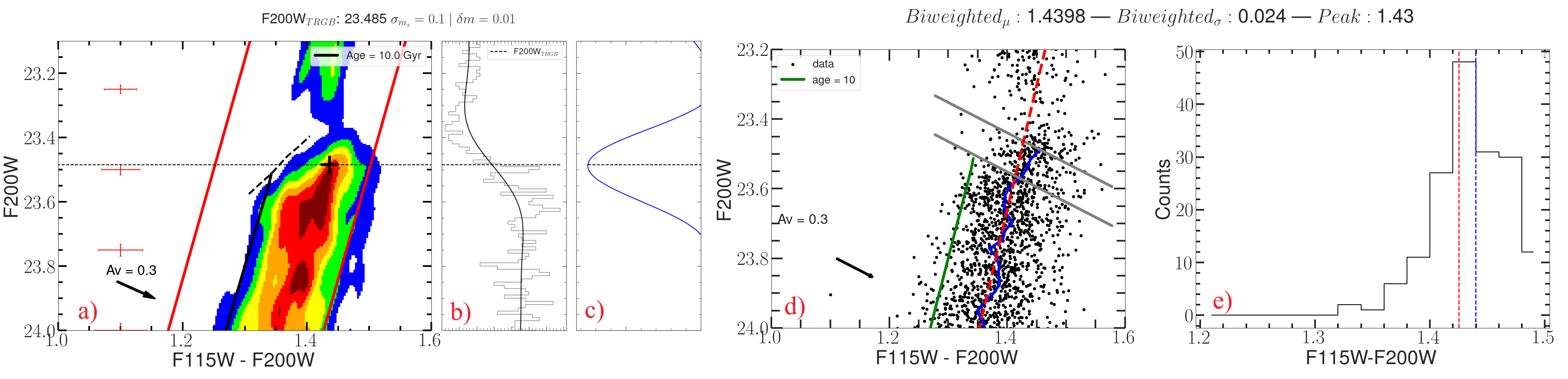}
\caption{The TRGB magnitude and color measurements using simulated CMD for two illustrative cases:  
(top) a 10~Gyr, Z=0.003 population with mean $A_V=0.1$~mag and $\sigma_{A_V}=0.2$~mag and (bottom) a population with  $9.8\leq\log{\rm(Age)}\leq10.0$, $0.002\leq Z\leq 0.003$ and $A_V=0.3$ and $\sigma_{A_V}=0.1$. (a) Illustration of TRGB magnitude measurement method on the simulated CMD. The colored contours indicate the kernel density of stars in the CMD. The locus of isochrones used for age-metallicity variations is indicated by black dashed line. (b) Binned F200W-LF (thin line) and its smoothed GLOESS component (thick line). (c) Sobel plot for the F200W LF. (d) The CMD corresponding to the kernel density diagram in (a) illustrating the TRGB color measurement. The parallel gray lines indicate the stars used for the bi-weighted color measurement. The blue curve indicates the median average of RGB stars' colors along the magnitude axis. The red dashed line indicates the line fitted to the blue curve. (e) The histogram of stars enclosed between the parallel lines indicated in (d). The blue and red dotted lines indicate the bi-weighted median and RGB-extrapolated colors, respectively.}
\label{fig:sim_Av_met}
\end{figure*}

\subsection{TRGB magnitude and color measurements on the simulated disk dataset}

We were unable to identify a sharp tip at the end of the RGB branch in Figure~\ref{fig:cmd}, which is mainly due to mixing stars coming from populations of distinct ages and metallicities that are experiencing distinct interstellar extinction. Such behavior is expected as the disks of galaxies are known to have formed stars over an extended period of time and presently contain gas and dust, leading to interstellar extinction. The problem can be simplified if we carry out the analysis over zones where variations in age, metallicity, and extinction are small, but the zones are still large enough to have a sufficient number of TRGB stars. By carrying out trials, we found that a square zone of 24~arcsec a side offers a good compromise. 
Such a region typically contains around 50--100 stars within 0.1~mag below the expected TRGB magnitude. We here first illustrate the technique of obtaining a TRGB magnitude in such a disk region, which is expected to be affected by non-negligible interstellar extinction and contamination from AGB stars. The recovered magnitudes and colors are also expected to be affected by crowding. We use simulated data for this illustration, which allows us to assess the accuracy of the recovered TRGB magnitudes and colors.

The simulations were carried out to mimic the conditions in the disk of NGC\,628, observed in JWST-NIRCam CMD. For this, we used PARSEC isochrones with ages $9.8\leq \log({\rm Age})\leq 10$ and metallicity $0.002\leq Z\leq 0.003$. We simulated four observational effects, namely, photometric errors, age-metallicity variation, extinction and its spread, and contamination from AGB stars. The observed photometric errors are modeled as Gaussian distributions in magnitude and color, with standard deviations in magnitude and colors corresponding to these in the observed data. The age-metallicity variation is chosen to mimic the $F115W-F200W$ color range in the 24~arcsec observed CMDs. The extinction is characterized by a Gaussian 
 distribution centered at $A_V$=0.3~mag, and a spread of ${A_V}$=0.1~mag. The AGB population is modeled as a uniform spread of stars along the evolutionary track corresponding to the AGB stage. In total, our simulations include $\sim$8000 RGB ($23.4\leq F200W\leq 24$) and $\sim$400 AGB stars, which represent conditions close to the central parts of the disk, where the AGB contamination affects the TRGB measurements, the most.

We show the effect of these physical phenomena in the four separate panels, respectively from left to right in Figure~\ref{fig:sim_CMD}. The simulations demonstrate that the RGB tip is flat in the absence of extinction and age-metallicity spread, with the width of the branch controlled by the 1-$\sigma$ error on color. The addition of an age-metallicity spread (panel b) introduces the well-known tilt in the RGB tip \citep[e.g.][]{trgb_jwst_theory, trgb_color_vs_z}. The addition of interstellar extinction (panel c) not only widens the branch but also changes the shape of the RGB tip. As expected, the RGB tip has an inclination along the reddening vector. The addition of AGB stars (panel d) introduces bright stars above the RGB branch.

The morphology of RGB stars in simulated CMDs provides key insights into the factors affecting observed RGB stars. Figure~\ref{fig:sim_Av_met} illustrates this, with the top panel showing TRGB color-magnitude measurements for 10~Gyr RGB stars ($Z=0.003$) with a mean extinction of 0.1 and an extinction spread of 0.2. The bottom panel presents TRGB color-magnitude measurements from the simulated CMD in Figure~\ref{fig:sim_CMD}, panel (d). 

\subsubsection{Measurement of TRGB magnitude}

We used the well-established edge detection technique \citep[e.g.][]{edge_det_1,edge_det_2,edge_det_3,edge_det_4,trgb_4258} for determining the TRGB magnitude in the F200W-band luminosity function (F200W-LF). To construct F200W-LF, we used stars along the theoretical isochrones with a spread of 0.2 magnitude. As the RGB branch is inclined in any CMD, the color-mag cut was performed parallel to the RGB branch in the F115W$-$F200W vs F200W CMD as illustrated in the panels (a) of Figure~\ref{fig:sim_Av_met}. The resulting F200W-LFs are shown in panels (b), where the continuous line is obtained using the Gaussian-windowed, Locally Weighted Scatter-plot Smoothing method (GLOESS) \citep{gloess_1, gloess_2}.
The edge-detection technique has been used successfully to quantify the TRGB magnitude in a reproducible way \citep{edge_det_1}. The panels (c) of 
Figure~\ref{fig:sim_Av_met} shows the results of the gradient search method for edge detection. The TRGB magnitude was defined as the peak found by the Sobel filter with kernel \{-1,0,1\}. We can see that the peak of the edge-detection coincides with the expected TRGB magnitude, which is the brightest point with the most number of RGB stars.

Unfortunately, the TRGB magnitude evaluated using this method depends on the bin sizes and the  $\sigma_s$ value of the Gaussian smoothing window used to construct the LF. We varied the bin sizes ($\Delta$m) in the range [0.001, 0.01] mag and $\sigma_s$ values in the range [0.05, 0.1] mag, finding a TRGB magnitude for each set of these values. We used the standard deviation in the resultant TRGB magnitudes as the error in the TRGB measurement and the median value as the TRGB magnitude.

\subsubsection{TRGB color measurement}
\label{sec:trgb_color}
For obtaining the TRGB color (F115W-F200W) we tested several approaches. As a first approach, we calculated the bi-weighted median color of stars within 5-sigma ($\sim$0.1 magnitude) of the measured TRGB magnitude. Specifically, we selected a magnitude window that covers 0.02 and 0.08 magnitudes above and below the TRGB magnitude, respectively. We performed iterative k-sigma clipping with $k=2$ using the bi-weighted median and bi-weighted standard deviation to estimate the TRGB color \citep[e.g.][]{trgb_color_vs_z}.  
In the second approach, we fitted a line to the RGB branch and extrapolated the fitted line (red line in Figure~\ref{fig:sim_Av_met}(d)) to meet the TRGB magnitude to obtain the TRGB color. The two measurements for the simulated data are shown in panels (e) of Figure~\ref{fig:sim_Av_met} along with the histogram of all stars used for the color analysis. By comparing the panels (e) in Figure~\ref{fig:sim_Av_met}, we infer that when extinction spread is dominant (top), the extrapolated RGB color is redder than the bi-weighted median color, whereas it is bluer when the age-metallicity spread is dominant (bottom). Comparing the input metallicity and extinction values to the retrieved ones, we find that the color from the first approach (bi-weighted median color) better fits the color of the TRGB stars.

The recovered values for the extinction-dominated case (top panel of Figure~\ref{fig:sim_Av_met}) are $Z=0.0033\pm0.0002$ and $A_V=0.05\pm0.03$, which agree well with the input values of $Z=0.003$ and $A_V=0.1$, considering that the Sobel filter picks the least extinct point. In the metallicity-spread-dominated case (bottom panel of Figure~\ref{fig:sim_Av_met}), the recovered values are $Z=0.0027\pm0.0002$ and $A_V=0.26\pm0.07$~mag, which agree well with the input values of $Z=0.003$ (metallicity at the brightest TRGB point) and mean extinction $A_V=0.3$ from the simulation in Figure~\ref{fig:sim_Av_met}. These results confirm the robustness of the bi-weighted median method for TRGB color measurements and the Sobel filter method for TRGB magnitude determination.

\begin{figure}
    \centering
    \includegraphics[width=\columnwidth]{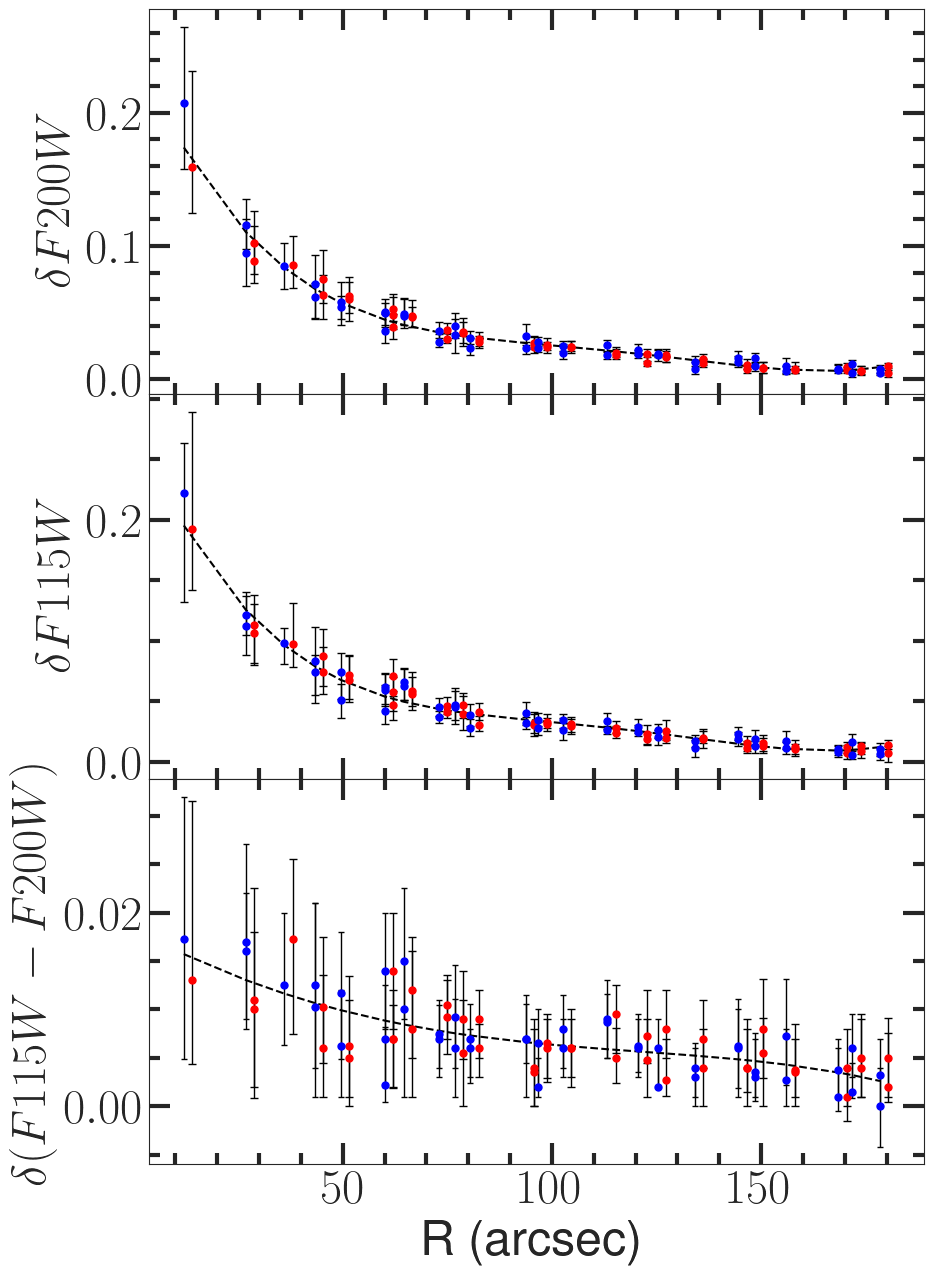}
    \caption{The radial dependence of the offset in the determined TRGB magnitudes and colors is due to the crowding as inferred from simulations. Blue and red bars correspond to regions in the North and Southern half, respectively, and the dotted line, a polynomial fit to the data. Each simulation consisted of recovering the individual magnitude of 100 artificial stars, and the median of these measurements. The experiment is repeated 100 times to get 100 median values, from which the input value is subtracted to get 100 offsets. The vertical bars at each position correspond to the 1-sigma value of these offsets, with the solid point on the line denoting the median offset.  }
    \label{fig:crowd_sim}
\end{figure}

\begin{figure*}
    \centering
\includegraphics[width=\textwidth]{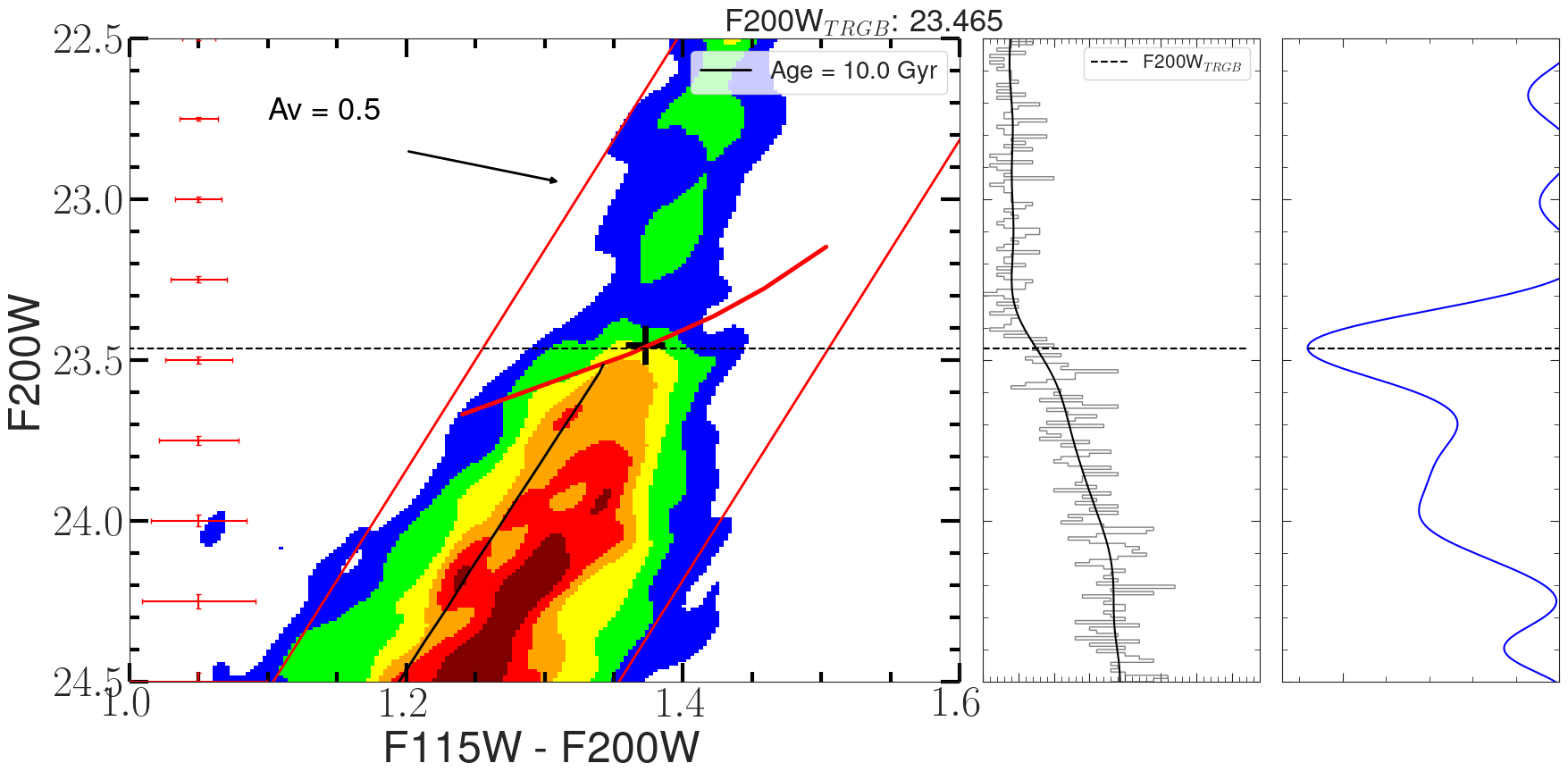}\\
    \caption{(Left) F115W$-$F200W vs. F200W kernel density diagram for an illustrative zone in NGC\,628 (region 10), where the red parallel lines indicate the color-magnitude range of stars used for constructing the F200W-LF. The black vertical line indicates 10~Gyr (Z=0.002) PARSEC isochrone. Typical error bars are shown by the red bars on the left at intervals of 0.25~mag. The reddening vector corresponding to $A_V$=0.5~mag is shown. The black "+" symbol marks the median TRGB point obtained from TRGB color-magnitude measurements. (Center) The binned (grey lines) and smoothed (black line) F200W-LF. (Right) The gradient of the F200W-LF was applied with a Sobel filter. The determined TRGB magnitude obtained from the edge-detection technique is written above the plot and is shown by the dashed horizontal line. Similar figures for all the 90 regions are available as a GIF file in the electronic version.}
    \label{fig:trgb_obs}
\end{figure*}

\begin{figure}
    \includegraphics[width=\columnwidth]{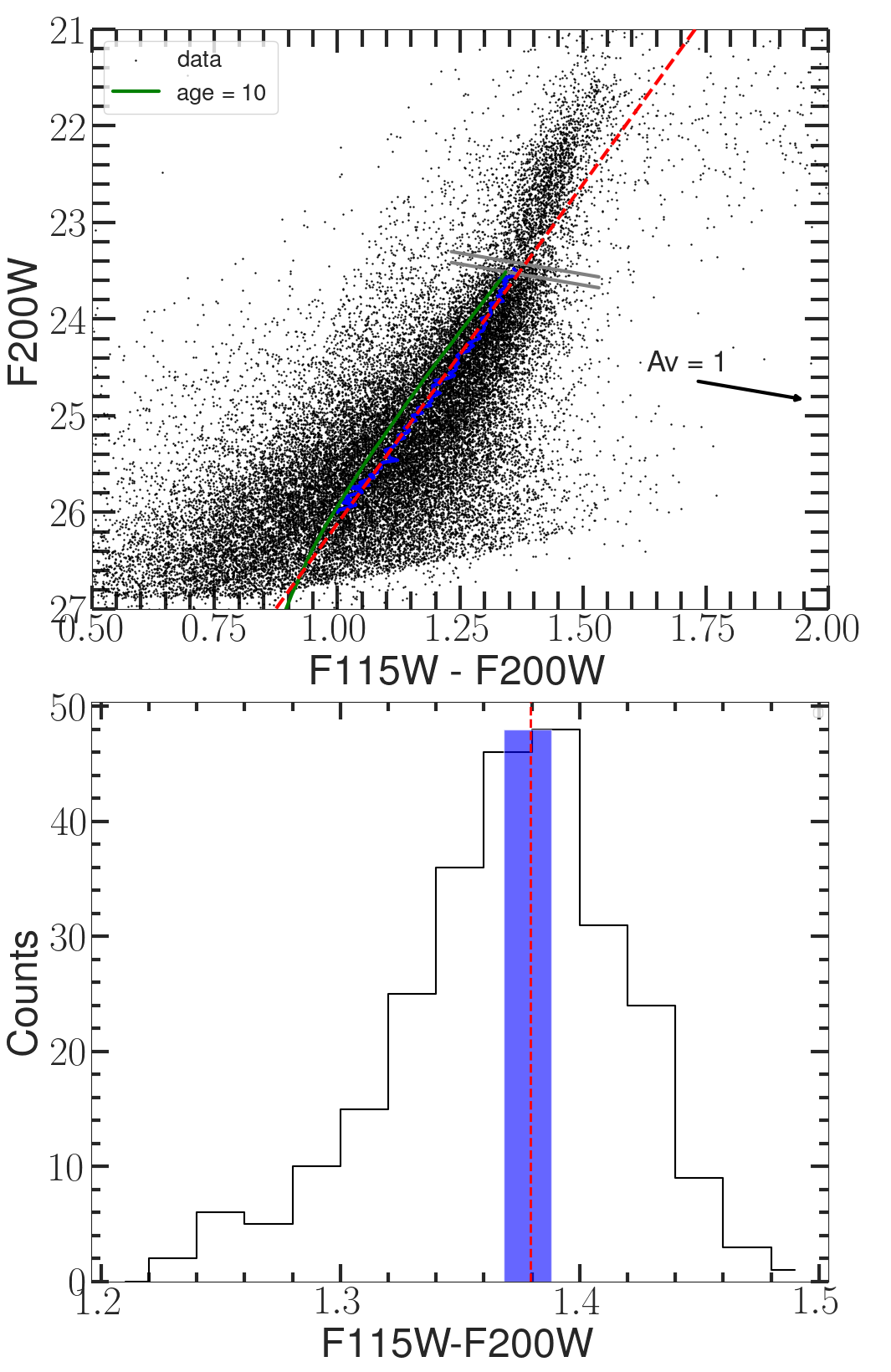}
     \caption{(Top) F115W$-$F200W vs. F200W CMD of region 10. The blue line indicates the median color found along the magnitude axis, fitting to the RGB. The red dashed line indicates the line fit to the blue line. The horizontal grey lines indicate the TRGB window with a magnitude width of 0.1~mag. The green line indicates 10~Gyr(Z=0.002) PARSEC isochrone. (Bottom) The color distribution of stars within the TRGB window for region 10, with the blue strip marking the bi-weighted mean (the blue line is broadened for better visibility as it overlaps with the red line), and the red dashed line representing the TRGB color obtained by extrapolating the RGB line fit.}
     \label{fig:trgb_color}
\end{figure}

\subsubsection{Effect of Crowding on TRGB color-magnitude measurements}

Crowding is a well-known factor that affects TRGB measurements, becoming particularly significant in disk regions. To evaluate its impact across the 90 analyzed regions, we revisited the completeness analysis by comparing simulated and retrieved magnitudes for the F115W and F200W filters. Unlike the completeness test, which involved $\sim$10,000 stars, we simulated 100 stars within each of the 90 zones to minimize the potential for introducing additional crowding and repeated the experiment 100 times, each time inserting stars in random positions. 

We specifically quantified crowding-induced shifts in magnitude and color, focusing on TRGB measurements. This was done by simulating stars with a magnitude of 25 in F115W and 23.5 in F200W, corresponding to an F115W$-$F200W color of 1.5. As shown in Figure~\ref{fig:crowd_sim}, crowding results in a radially-dependent color-magnitude offset, which affects the CMDs and subsequently the TRGB measurements.
The radial dependence on either side of the center (blue and red) almost coincides. The sense of the offset is such that the observed magnitudes are brighter. The offsets, as well as errors on them, are systematically higher in the crowded inner regions. The observationally obtained TRGB measurements using the edge-detection technique are the median magnitude of the population, rather than that of a single star. This implies that we can recover the crowding-corrected magnitudes after applying the median offsets obtained from the simulations for each region, without additional errors due to crowding. Hence the median difference between input and output magnitudes and colors in these simulations was adopted as the offset due to crowding. The variation in the medians across the 100 trials was used to estimate the error in this offset. These offsets were applied to all TRGB color-magnitude measurements, and the associated errors were incorporated into the overall TRGB measurement uncertainties.

\subsection{TRGB color-magnitude measurements in the disk of NGC 628}

We followed the same technique as for the simulated data to obtain TRGB magnitude and color for each region in the disk. In Figure~\ref{fig:trgb_obs}, we show the procedure for an illustrative case (region 10), where three panels from left to right show the Kernel-density plot of the CMD, F200W-LF, and the Sobel plot, respectively, where all the notations are similar to those in Figure~\ref{fig:sim_Av_met}(a,b,c) for the simulated data. The Sobel peak marking the TRGB is well-defined. We have used a segmented color-map to show the shape of the density gradient around the TRGB. The inclined shape (yellow contours) very much resembles the simulated case in the presence of extinction, and is not along the direction expected for a metallicity gradient (see Figure~\ref{fig:sim_CMD}(b)).

Figure~\ref{fig:trgb_color} (top) shows the CMD for the illustrative region 10. As compared to the simulated CMD (Figure~\ref{fig:sim_CMD}(c)), the RGB is broader, with the extinction responsible for the broadening on the red side, whereas intermediate-age disk stars occupy the bluer side. To obtain the TRGB color, we extrapolated the RGB slope (red-dashed line) to meet the TRGB magnitude obtained from the Sobel method. In the bottom panel, we show the color histogram of stars in the TRGB window (gray parallel lines), where the red vertical line denotes the extrapolated RGB color. This color agrees very well with the color obtained from the bi-weighted median (the center of the broad blue strip).

\subsection{Spatially resolved TRGB magnitude and color measurements}\label{sec:spatial_trgb}

\begin{figure*}
    \centering
 \includegraphics[width=\textwidth,trim={4.5cm 2cm 4.4cm 0cm},clip]{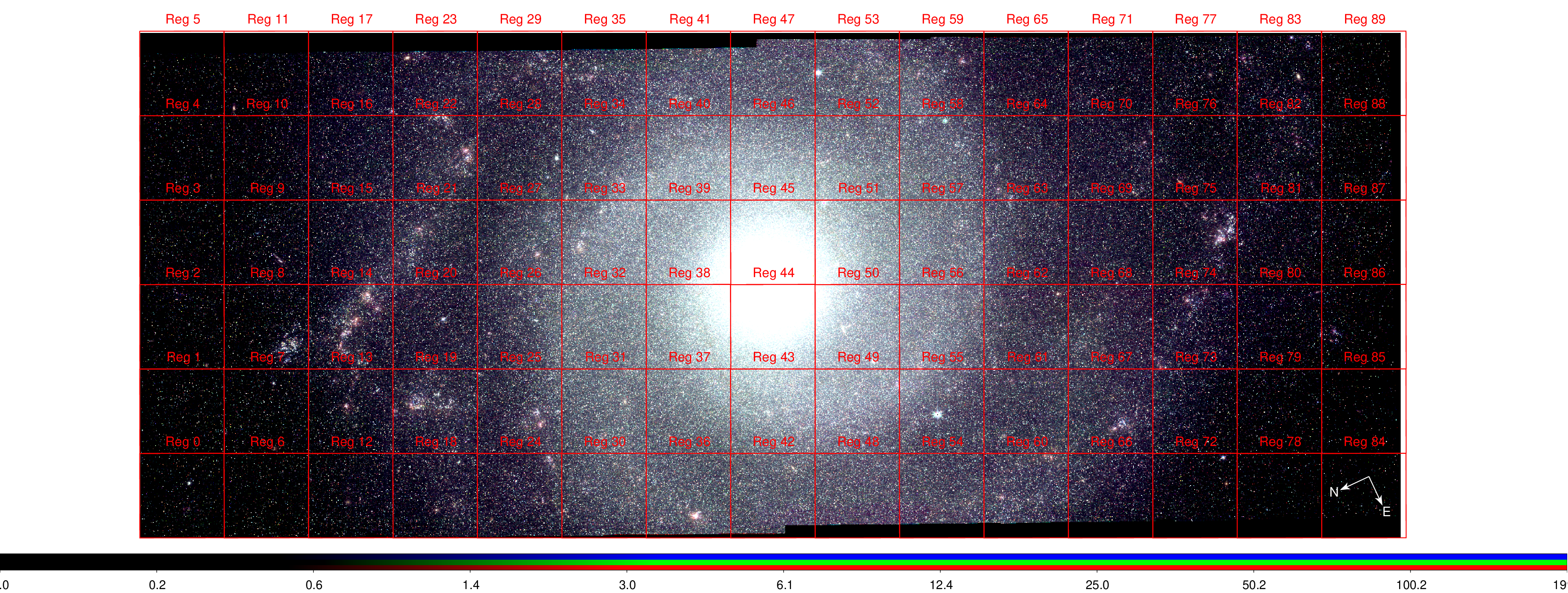}
    \caption{The 24~arcsec$\times$24~arcsec zones where we carried out the CMD analysis to obtain the F200W TRGB magnitude and its F115W$-$F200W color is shown overlaid on a NIRCam RGB image formed using F200W, F150W, and F115W as red, green and blue components, respectively. The image is oriented along the detector XY axes. The central six zones are discarded from the analysis because the expected TRGB magnitude is only marginally brighter than the completeness magnitude in the bulge regions, as illustrated in the top panel of Figure~\ref{fig:crowd_corr}.}
    \label{fig:nircam_rgb}
\end{figure*}

We illustrated above that zones of 24~arcsec a side have enough stars for identifying the TRGB point using edge detection technique in the F115W$-$F200W vs F200W CMD.
Motivated by this, we segmented the entire NIRCam FoV into square regions measuring 24~arcsec in size (see Figure~\ref{fig:nircam_rgb}). This segmentation enables us to leverage TRGB measurements for spatially resolved estimates of stellar metallicity and extinction. The size of 24~arcsec a side is the optimized value obtained after extensive trials to balance between spatial resolution and the error on the determined TRGB magnitude. The larger the number of stars used in the analysis, the smaller the error on the TRGB magnitude obtained using the edge detection method. Although sampling larger areas gives us a larger number of stars, it also introduces spread in the TRGB magnitude and colors due to the star-formation history and metallicity variations in the disk. For an analysis area of 24~arcsec side square, we get approximately 8000 stars, of which 50--200 belong to the tip of the F200W-LF, with a median value of 100 TRGB stars, which ensures that the tip is sufficiently populated \citep{trgb_jwst_theory}. Here, the tip corresponds to the magnitude window defined in the TRGB color-measurement section (Section~\ref{sec:trgb_color}), with a color window of 0.1 magnitudes in width centered on the estimated TRGB color.

\begin{figure}
    \centering
    \includegraphics[width=\columnwidth]
    {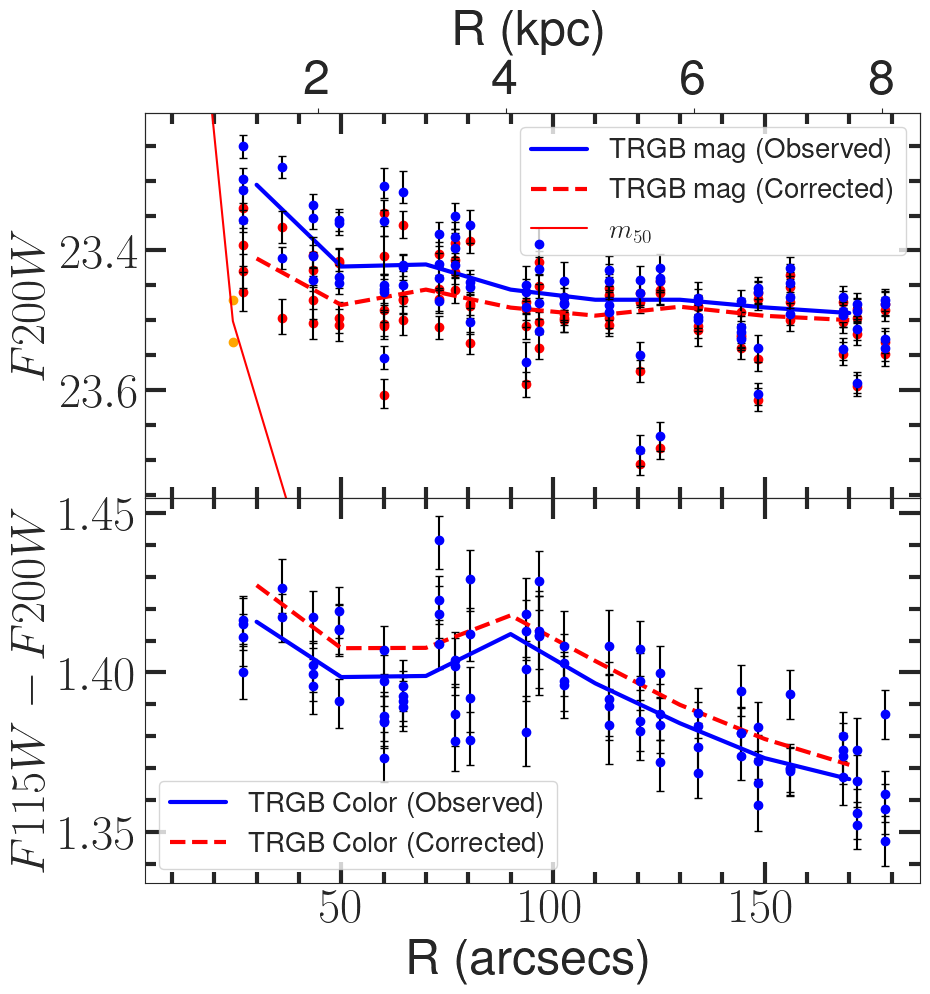}
    \caption{(Top) Blue circles indicate measured TRGB magnitudes as a function of galactocentric radius and red circles indicate TRGB magnitude measurements rectified for crowding. The red solid line corresponds to $m_{50}$ magnitude as a function of galactocentric radius (Bottom) $m_{50}$ as a function of galactocentric radius for select 15 regions along the radius.}
    \label{fig:crowd_corr}
\end{figure}

\begin{figure}
    \centering
\includegraphics[width=\columnwidth]{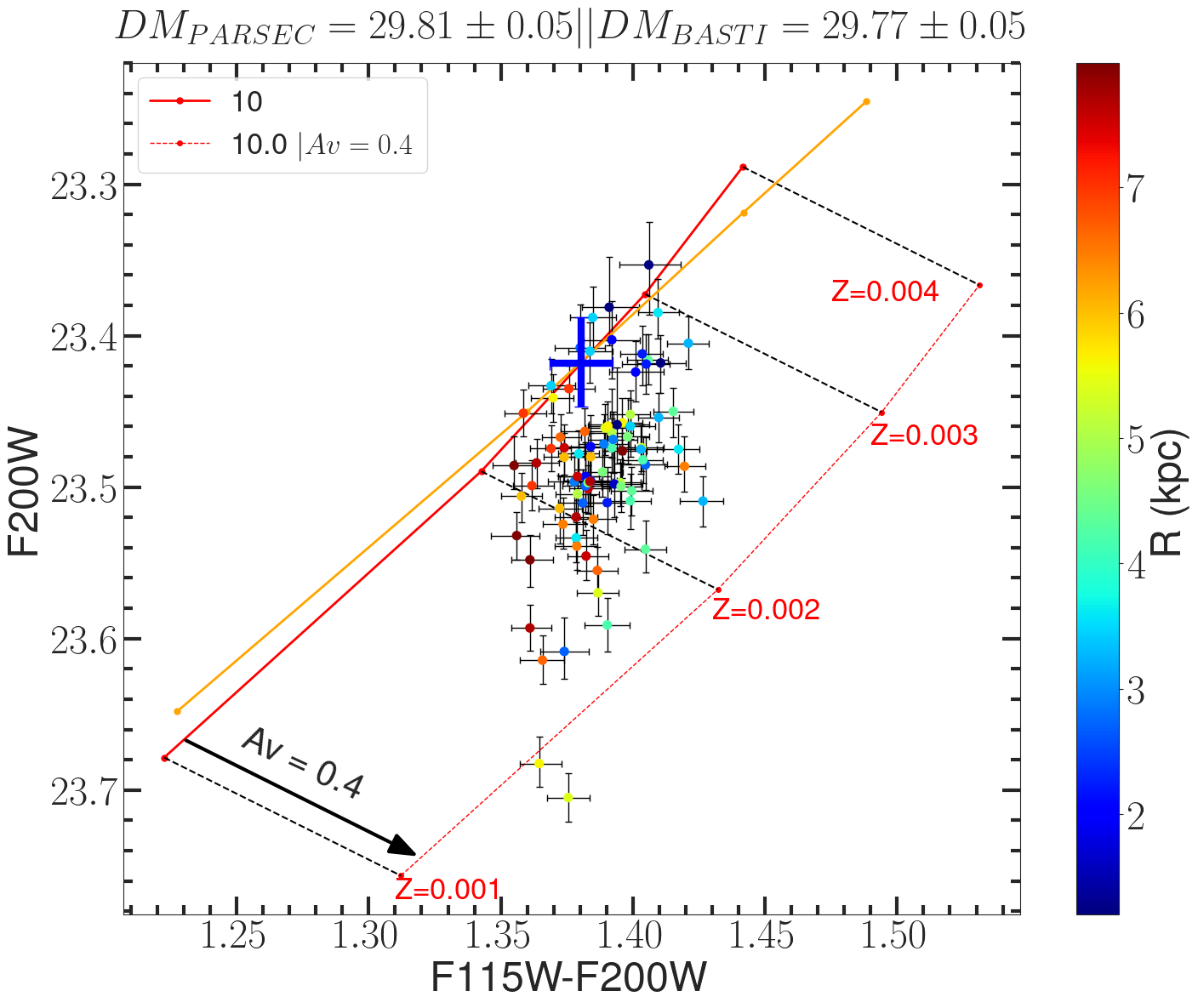}%
    \caption{TRGB magnitude vs. TRGB color for the 90 zones (filled circles color-coded based on their distance to the galaxy center following the color bar to the right). The thick red and orange lines join the theoretical TRGB magnitudes and colors from PARSEC and BaSTI isochrones, respectively, at 10~Gyr at metallicities between Z=0.001 to 0.004. The dotted red line shows the 10~Gyr PARSEC line shifted by a reddening equivalent to $A_V$=0.4~mag. The blue cross corresponds to TRGB measurements of the super-bubble Section~\ref{subsection:TRGB_distance}). The theoretical TRGB absolute magnitudes are brought to the observed magnitude using the DM obtained for this bubble using $A_V$=0~mag.
    }
     
    \label{fig:dismod}
\end{figure}
\begin{figure*}
    \centering
\includegraphics[width=0.7\columnwidth]{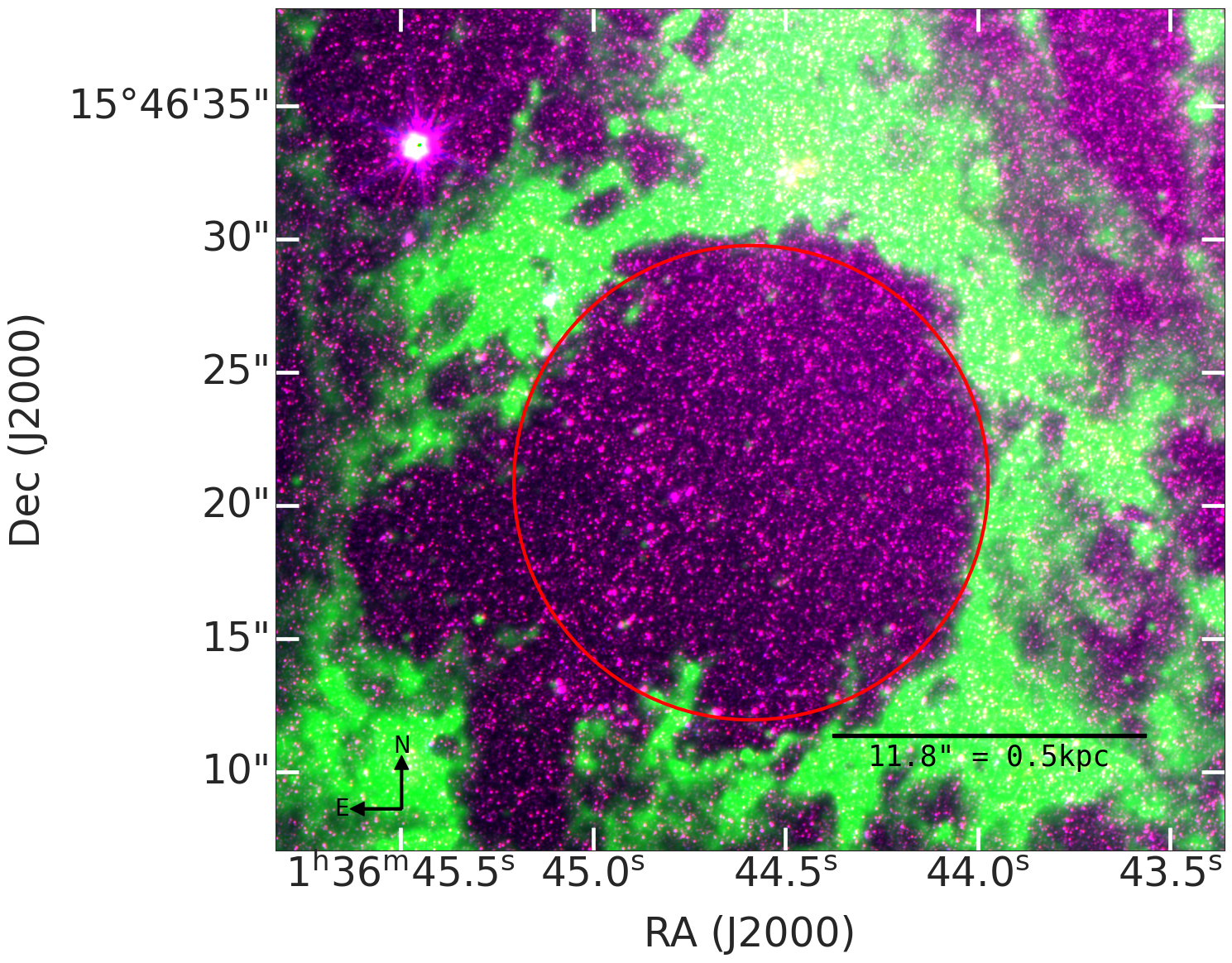}
\includegraphics[width=0.62\textwidth]{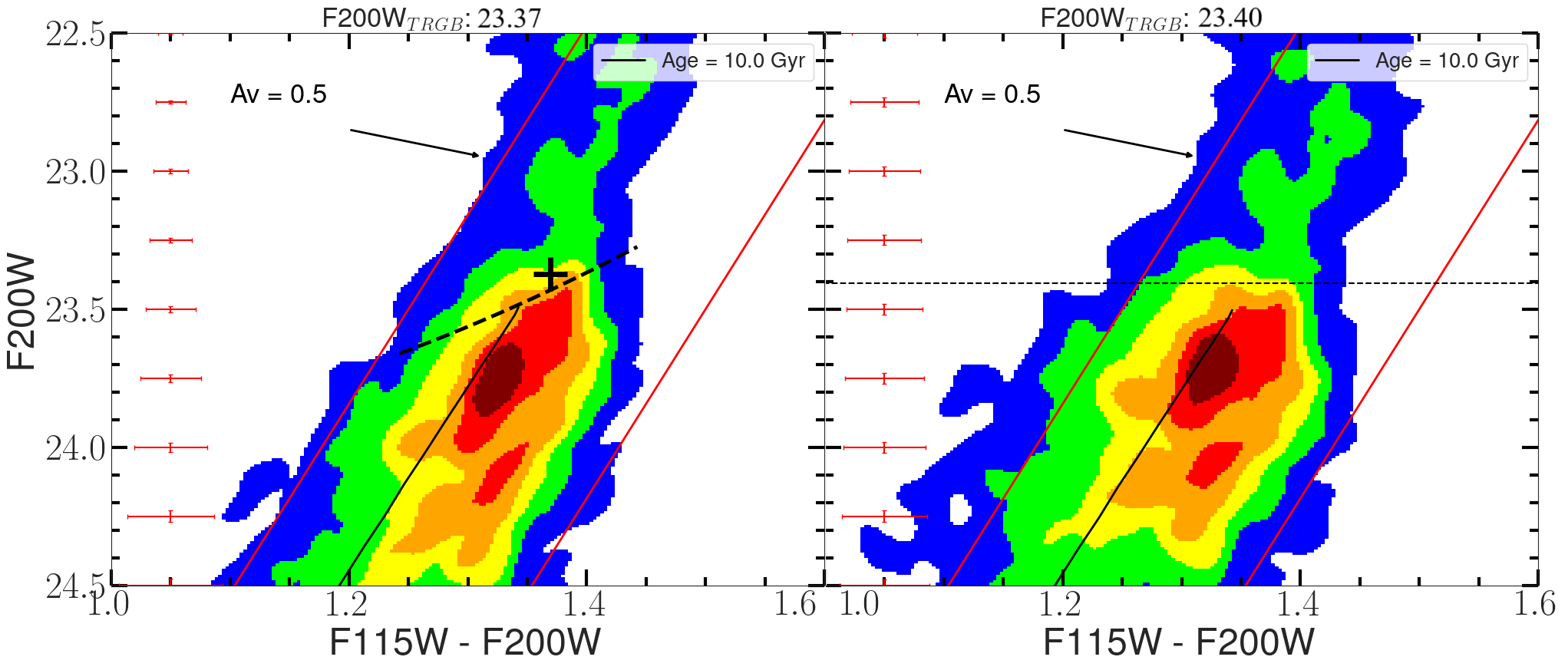}
    \caption{Left: RGB image of super-bubble in NGC\,628 formed using F200W, F770W, and F115W filter images as the red, green, and blue components, respectively. The red circle indicates the 8.9~arcsec region utilized for TRGB magnitude estimation. Middle: Same as Figure~\ref{fig:trgb_obs} (left), but for the stars seen through the super bubble shown on the left. The black thick dashed curve shows the locus of PARSEC isochrones with 4~Gyr$\leq$ Age$\leq$13~Gyr and 0.001$\leq Z\leq 0.006$ Right: Same as the middle panel but the CMD is corrected for RGB-tip tilt due to age-metallicity variation.}
    \label{fig:superbubble} 
\end{figure*}

The analysis of the illustrative region 10 gave us  F200W$_{TRGB}$=23.453$\pm{0.016}$~mag and $F115W-F200W_{TRGB}$=1.37$\pm0.009$~mag. We obtained TRGB magnitudes and colors for the 90 regions, which are shown in Figure~\ref{fig:crowd_corr} as a function of the galactocentric distance of the region (blue points with vertical error bars). As we illustrated in Figure~\ref{fig:crowd_sim}, the brightness is overestimated due to crowding with a clear radial dependence. We corrected each magnitude and color due to the effect of crowding and showed the resulting values as red dots (we do not show the red dots in the color diagram as the correction is too small). The solid blue and red dashed lines show the average radial dependence, respectively, before and after the correction for crowding. Even after the correction, a residual gradient is noticeable in both magnitude and color, with the TRGBs marginally brighter, and redder towards the center. Large dispersions are also seen at a given galactocentric distance.

In addition to the measurement errors introduced by the edge detection method, TRGB measurements also suffer from errors due to selection effects, contamination from reddened RSG stars, TP-AGB stars, zero-point errors, errors introduced by the photometry routine (DOLPHOT), etc. For accounting for the effects of photometric errors, we perform Monte-Carlo (MC) simulations by sampling photometric magnitudes for each star from a Gaussian distribution with a $\sigma$ corresponding to the photometric error and $\mu$ corresponding to the observed magnitude, under different sets of smoothing and binning parameters. From this analysis, we obtained errors of $\sim$0.01 and $\sim$0.006 in TRGB magnitude and color, respectively, and $\sim$0.02 and $\sim$0.009 for regions within 70 arcseconds. In  Table~\ref{tab:systematic_error}, we summarize the statistical and systematic uncertainty budget on TRGB color-magnitude and DM measurements presented in this work.

\begin{deluxetable}{lcc}
\tablecaption{Uncertainty budget on TRGB color-magnitude measurements. The errors given are based on measurements on stars projected inside the Superbubble.}
\tablewidth{0pt}
\colnumbers
\tablehead{\colhead{Uncertainty}&\colhead{$\sigma_{\rm stat}\,[mag]$}& \colhead{$\sigma_{\rm sys}\,[mag]$}  \\
\colhead{} & \colhead{(F200W, F115W-F200W)} & \colhead{} }%
\startdata  
Photometric          & 0.02, 0.004 &     -    \\
Binning \& Smoothing & 0.03, 0.008 &     -    \\
Crowding             & 0.01, 0.008 &     -    \\
Isochrone Model      & -           & 0.04     \\ 
Age+metallicity spread &     -     & 0.03     \\
PSF stability        & -           & 0.01\textsuperscript{a} \\
DOLPHOT Photometry   & -           & 0.03\textsuperscript{a} \\
NIRCam Zeropoints    & -           & 0.02\textsuperscript{a} \\
\hline
$A_{\rm F115W}/A_{\rm V}$      & -           & 0.05\textsuperscript{b} \\
$A_{\rm F200W}/A_{\rm V}$      & -           & 0.04\textsuperscript{a} \\
Foreground $A_{\rm V}$     & -           & 0.004\textsuperscript{c}\\
\hline
Total Reddening error &   -  & 0.008\textsuperscript{d} \\
Pivot color selection & 0.03, -    & -    \\ 
\hline
Distance Modulus    & 0.05        & 0.063\\
\enddata
\tablecomments{$a$ Obtained from \cite{trgb_4258}.\\ $b$ \cite{av_law}\\ $c$ \dataset[ADS/IRSA.Dust\#2025/0301/130524\_7577]{https://irsa.ipac.caltech.edu/applications/DataTag/}\\
$d$ $(A_{\rm V} + \sigma_{A_{\rm V}})\times (A_{\rm F200W}/A_{\rm V} + \sigma_{(A_{\rm F200W}/A_{\rm V})} )  - (A_{\rm F200W}/A_{\rm V})\times A_{\rm V}$}
\label{tab:systematic_error}
\end{deluxetable}

The increased error in the central regions is primarily attributed to a combination of crowding, AGB contamination, and effects from the star formation history. The cumulative error from the MC simulations, which accounts for binning, smoothing, and photometric uncertainties, is combined in quadrature with the crowding error to derive the total error in TRGB magnitude and color measurements. The median error in the TRGB mag and color measurements are  0.017~mag, and 0.008~mag, respectively. Hence, the measurement error is not the cause of the vertical spread of the points at a given radius. From Table~\ref{tab:trgb} (and also Figure~\ref{fig:parsec_iso}), we know that the F200W TRGB magnitudes are very sensitive to metallicity and $A_V$, with a weaker dependence on age. In the following section, we analyze the magnitude and colors of TRGB measurements in the 90 zones to determine the physical quantities across the galaxy's disk.

\section{Results \& Discussion}
\label{sec:result}

We now analyze the derived TRGB magnitudes and colors of all 90 regions to explore the possibilities of determining useful physical quantities such as metallicity and $A_V$ of the disk populations in each region. In Figure~\ref{fig:dismod}, we plot all the TRGB points in the F115W$-$F200W vs F200W CMD. This figure is similar to the right panel of Figure~\ref{fig:parsec_iso}, except that this figure contains observed points, and only the 10~Gyr TRGB locus as a function of metallicity is shown. As illustrated before, the TRGB magnitudes after correction for the offset due to crowding span a range of 0.20~mag. In comparison, the color varies by  0.10~mag. Significantly, the points are not scattered all over, instead brighter TRGB magnitudes have redder colors and the fainter TRGB magnitudes have bluer colors, which is not expected if extinction is responsible for the observed spread. 

To facilitate the interpretation of the observed trend of points, we also plot the grid showing the dependencies of the determined TRGB on metallicity and extinction. The superposition of theoretical isochrones requires the use of an internally consistent DM to the galaxy, which we found to be 29.81~mag using PARSEC models as discussed in the next subsection.

\subsection{JWST TRGB distance to NGC 628}
\label{subsection:TRGB_distance}

JWST observations of nearby star-forming galaxies using the MIRI filters have revealed the presence of numerous bubbles in the disks of these galaxies \citep{phangs}. NGC\,628 hosts one of the largest bubbles with a diameter of 1~kpc, which \citet{Barnes_2023} referred to as the Phantom Void. \citet{Mayya_2023} established that this void is in fact a bubble formed due to the feedback from massive stars. They found almost no detectable emission in any of the interstellar tracers, suggesting the bubble is empty of gas and dust. Though the bubble has recently formed stars, these stars account for only about 10\% of the total stars in the resolved stellar population, with the rest of the stars belonging to the disk population. This makes the bubble an excellent zone for the determination of the TRGB magnitude with possibly the least affectation from interstellar extinction. A color-composite RGB image of this bubble in JWST filters was carefully selected to depict the bubble as well as the disk stars shown in the left-most part of Figure~\ref{fig:superbubble}. 

We constructed a CMD using stars that are seen projected onto the empty region of the bubble. For this, we used all stars that are within a radius of 8.9~arcsec from the bubble center (the red circle in Figure~\ref{fig:superbubble}). Assuming minimal extinction effects on the super-bubble, we can apply the standard TRGB measurement technique to its RGB stars for distance estimation. This involves correcting the RGB-tip tilt caused by age-metallicity variations using theoretical or empirical curves and detecting the TRGB magnitude with the Sobel method.  The TRGB magnitude of the stars in the bubble is illustrated in Figure~\ref{fig:superbubble} (b). We tested both PARSEC (Figure~\ref{fig:superbubble} (b)) and BaSTI \citep{basti} models for age-metallicity correction. The edge-detection technique clearly detects a peak corresponding to the TRGB point at F200W$_{TRGB}=23.40\pm0.05$~mag. We estimated crowding offset at the bubble location with the same strategy as the 90 zones, which corresponds to $\sim0.05 \pm 0.01$~mag and $\sim0.01 \pm 0.008 $~mag in F200W and $F115W-F200W$, respectively. The tilt-corrected TRGB-magnitude measurement yielded distance moduli of $29.80\pm0.05$~mag and $29.83\pm0.05$~mag using PARSEC and BaSTI models, respectively.

We also carried out an analysis of the CMD without tilt correction to determine the TRGB magnitude and color, yielding  $23.37\pm0.03$~mag and $1.38\pm0.01$, respectively. The magnitude and color of the bubble TRGB stars after correcting for the crowding offset are plotted as a blue cross in Figure~\ref{fig:dismod}. We determined an internally consistent DM in such a way that the line joining the TRGB point at different metallicities for 10~Gyr isochrones (solid red and orange lines for PARSEC and BaSTI, respectively) passes through this point in this figure, which makes the hypothesis that stars seen projected onto the bubble have $A_V$=0~mag. TRGB magnitude and color measurements gave 29.81$\pm$0.05 (stat)~mag with PARSEC and 29.77$\pm$0.05~(stat)~mag with BaSTI, showing close agreement between methods. The TRGB point is very well defined for the bubble despite having a smaller area compared to the 24~arcsec grid, which indicates no spread in the TRGB point due to reddening. The determined TRGB mag and color indicate Z=0.0027 with PARSEC and Z=0.0023 with BaSTI for a TRGB age of 10~Gyr. Finally, we use the DM of 29.81~mag from the TRGB color-magnitude measurements because the tilt-corrected TRGB measurement has a systematic dependence on age-metallicity models used for tilt correction, although within the error. The DM of 29.81$\pm$0.05~mag corresponds to a distance of 9.25$\pm$0.4~(stat)~Mpc.

Our newly obtained DM is consistent with the DM=$29.89^{+0.06}_{-0.09}$~mag obtained using PNLF by \citet{pnlf1}, who also assumed Av=0~mag for the brightest PNe. On the other hand, the TRGB distance moduli obtained from HST in different studies cover a wide range from 29.68 to 30.04 \citep{trgb_hst_1,trgb_hst_2, trgb_hst_3}, with our values lying in the middle of this range.

\subsection{Spatially resolved extinction and stellar metallicities}
\begin{figure}[!htbp]
    \centering
\includegraphics[width=\columnwidth]{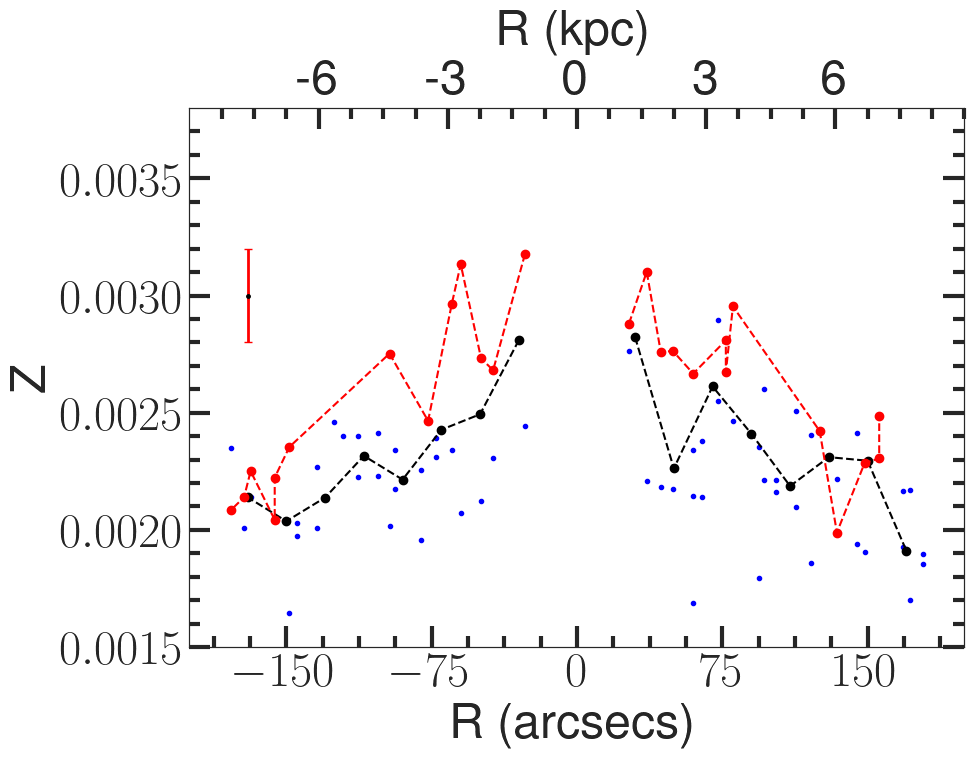}\\
\includegraphics[width=\columnwidth]{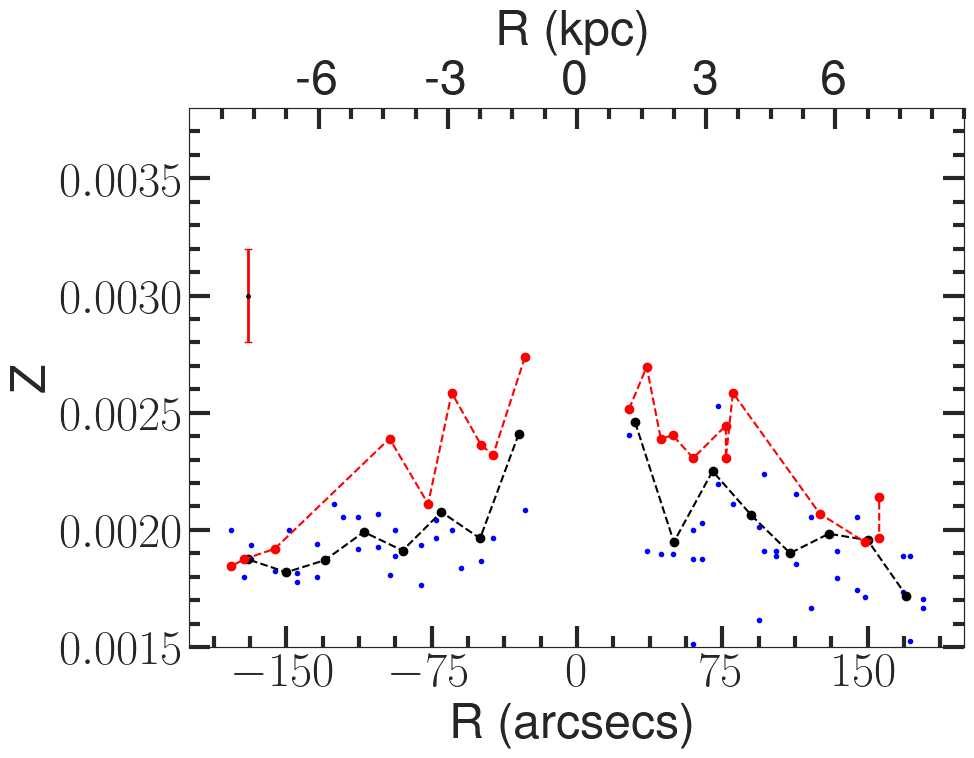}
    \caption{Stellar metallicity derived from analyzing the TRGB colors and magnitudes, plotted against distance from the center. A negative radius indicates a negative x-axis with the origin at the galaxy's center (the x and y axes parallel the pixel axes). The black line indicates radially binned stellar metallicity values. Red dots connected by red dashed line indicate metallicity measurements from low extinction ($A_V<0.06$~mag) regions. (Top) Stellar metallicity was obtained using the PARSEC isochrones. (Bottom) Stellar metallicity was obtained using the BaSTI isochrones. }
    \label{fig:Z_star}
\end{figure}

\begin{figure}[!htbp]
    \centering
\includegraphics[width=\columnwidth]{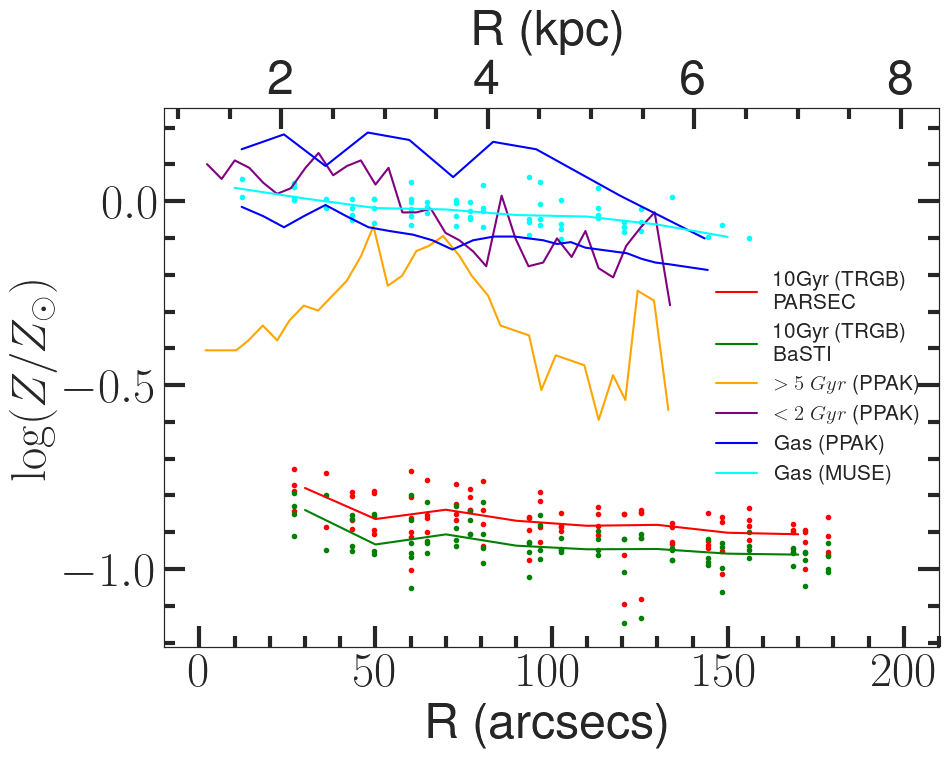}
    \caption{Metallicity Gradients using tracers of different ages.  Red circles represent stellar metallicity estimates derived from TRGB measurements using PARSEC isochrones, with the red line joining the median values at selected radii. Similarly, green circles represent stellar metallicity estimates derived from TRGB measurements using BaSTI isochrones, with the green line joining the median values at selected radii. Cyan circles depict gas-phase metallicity estimates obtained using the N2 metallicity relation from PHANGS MUSE emission line maps. The median of these points is shown by the cyan line. The dark blue line represents the gas-phase metallicity gradient obtained by \citet{ppak} using two different calibrators: $ff-T_{eff}$ and O3N2. Stellar metallicities for two populations of ages $<$2~Gyr(purple) and $>$5~Gyr(orange) from \citet{disk_age}.}
    \label{fig:z_grad}
\end{figure}

\begin{figure}[!htbp]
    \centering
\includegraphics[width=\columnwidth]{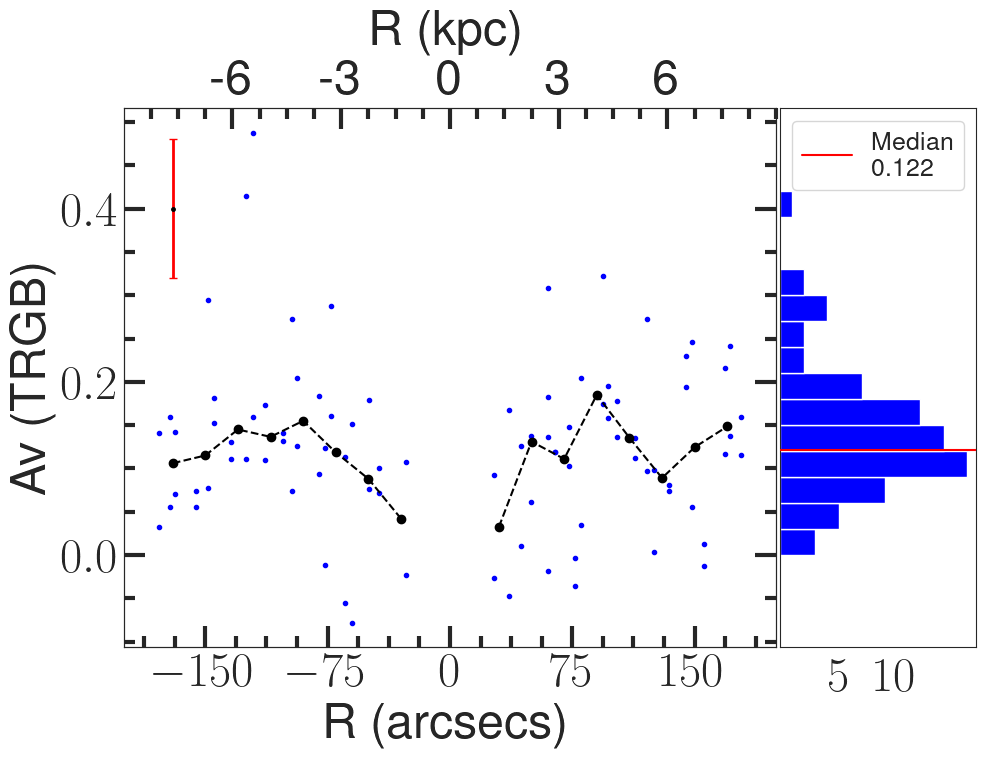}
\includegraphics[width=\columnwidth]{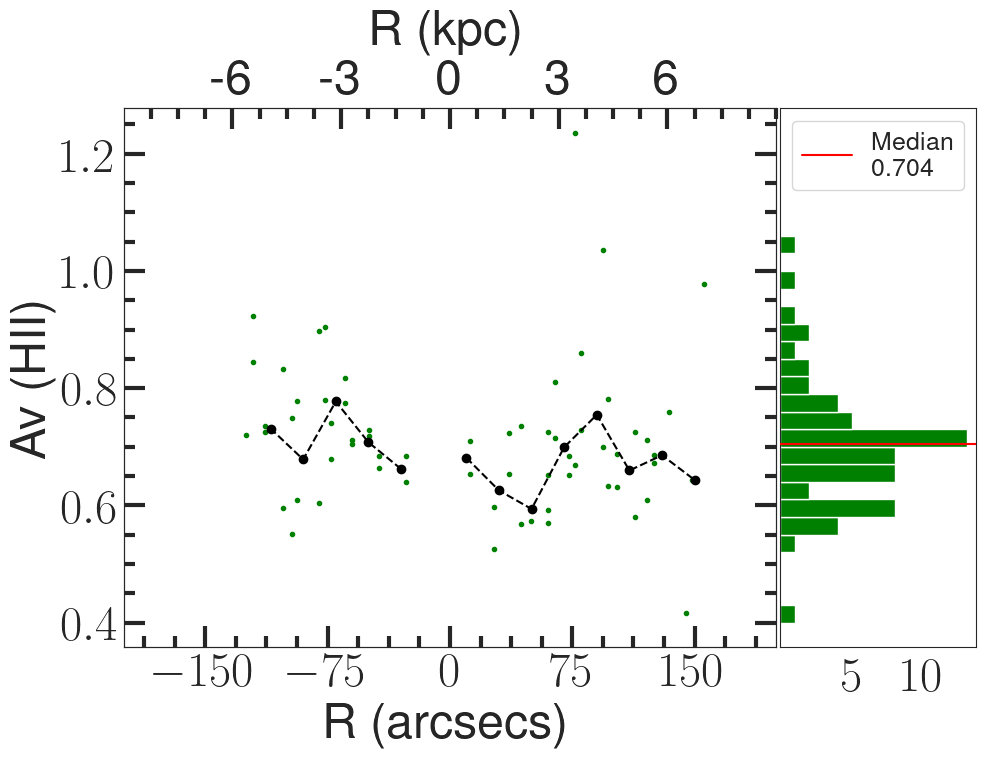}
    \caption{The top panel displays extinction values, represented
by blue-filled circles, estimated using TRGB measurements,
with the black line indicating radially binned values. The bottom panel shows gas extinction, represented by green-filled circles, obtained using the Balmer decrement from PHANGS MUSE emission line maps. Histograms of both the determined $A_V$ values are shown in the right panels, with the median of the distribution shown by the red line.}
    \label{fig:Av}
\end{figure}

Having obtained a new DM consistent with our measured TRGB points, we now use the TRGB magnitude and color to extract the extinction and stellar metallicities. 

We determined each region's metallicity and extinction by moving the TRGB points to the theoretical line along the reddening vector. The length of the vector represents the extinction value, while the intersecting point provides the metallicity for a given age. We use the 10~Gyr isochrone following the study of \citet{disk_age}, which found populations of this age throughout the disk of NGC\,628.

The metallicity values determined for the 90 zones are plotted against the mean distance from the galaxy center in Figure~\ref{fig:Z_star} using PARSEC and BaSTI isochrones in the top and bottom panels, respectively. Median values in radial bins are connected by black dashed line, which illustrates that the TRGB metallicity systematically decreases with increasing galactocentric distance with the profile behaving similarly on opposite sides of the center, for both sets of evolutionary models. However, there is an offset of 0.0002 ($\sim$10\%), with the BaSTI models giving systematically lower values. The dispersion of metallicity at each radial bin is slightly larger than the measurement errors. In order to understand this dispersion, we show the points with lowest extinctions ($A_V<$0.06~mag) by red dots which are joined by red dashed lines. All these points mark the highest metallicities at a given radius. As pointed out earlier, the Sobel method to determine the TRGB magnitude picks the most metal-rich TRGBs at low extinction regions, and hence it is expected that the low $A_V$ regions have systematically higher metallicity. At higher extinctions, the measured TRGB magnitudes correspond to lower metallicities, implying the presence of an intrinsic dispersion in metallicity in each measured zone.

In Figure~\ref{fig:z_grad}, we compare the metallicity gradient obtained from TRGBs with the gas-phase and stellar metallicities reported in the literature. The gas-phase metallicities are obtained using H\,{\sc ii} regions by Rosales-Ortega et al. (2011), which represents the current values, whereas the TRGB metallicity corresponds to the metallicity prevalent $\sim$10~Gyr ago when the disk was young. In order to understand the large difference in metallicities of the youngest and oldest components of the disk, we plot the stellar metallicities at two intermediate ages provided by \citet{disk_age}. The metallicities of the populations younger than 2~Gyr are in general close to the current-day gas metallicities, whereas the metallicities of the populations older than $>$5~Gyr age are intermediate between the TRGB metallicities and populations younger than 2~Gyr. This illustrates a gradual build-up of metallicities from the 10~Gyr to the current epochs.

We now plot the $A_V$ values determined for the 90 zones against the mean distance of the zone from the galaxy center in Figure~\ref{fig:Av}. The extinction values were found to be almost model-independent. Figure~\ref{fig:Av} shows extinction values obtained using PARSEC isochrones.
The trend in the extinction behavior with radius is not as clear as that for the metallicity. The TRGBs are old stars that are uniformly distributed in the disk, and hence the derived $A_V$ values should be coming from the dust distributed in the diffuse component. The measurements in the central zones are consistent with zero extinctions, with a trend for $A_V$ to increase towards intermediate zones reaching a maximum value of 0.25~mag. The recent analysis of the central region of NGC\,628 using JWST data suggests a lack of PAH features. The H{\sc i} \citep{THINGS} and H$\alpha$ maps also suggest a lack of gas in the central regions. Hence, the low values of $A_V$ we obtained for the central regions are consistent with the gas-poor nature of this region.
The disk regions in the radial zone between 50--150~arcsec have $A_V>0.1$~mag, with a median value of 0.12~mag. Beyond 150~arcsec extinction appears to be reducing.

In the bottom panel of Figure~\ref{fig:Av}, we plot the gas extinction obtained from Balmer decrement ratios using PHANGS emission line maps generated from MUSE data. The Balmer decrement ratio is the median of all spaxels with the H$\beta$ S/N $>$3 within each 24~arcsec region. The $A_V$ measured using H{\sc ii} regions sample the conditions close to star-forming regions, such as in the spiral arm of the galaxy. Such regions are expected to contain dense clumps of gas and dust that would offer more extinction than that offered by diffuse interstellar dust. Thus, $A_V$(H{\sc ii}) is expected to be higher than $A_V$(TRGB). We find an almost uniform $A_V$(H{\sc ii})=0.70$\pm$0.1~mag throughout the disk.

\subsection{Inferences from TRGB color measurements}

\begin{figure}
    \centering
    \includegraphics[width=\columnwidth]{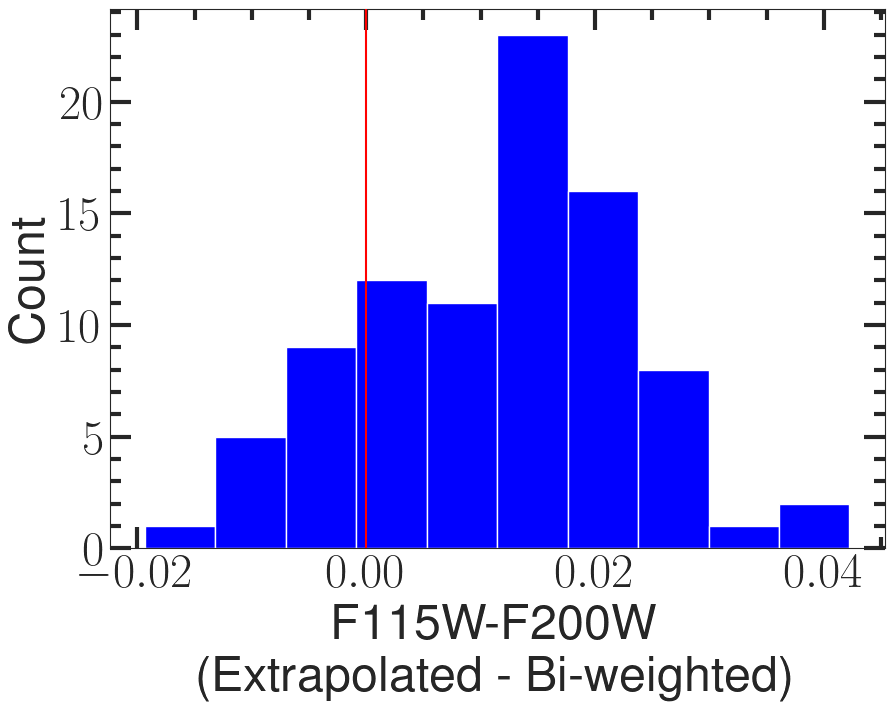}
    \caption{Distribution of the difference between extrapolated RGB color and bi-weighted median color in TRGB measurements.}
    \label{fig:col_diff}
\end{figure}

In this work, we have demonstrated that it is possible to obtain spatially resolved TRGB mag measurements in the disks of galaxies using CMDs uncorrected for the tilt of the RGB tip. The measured metallicities and their spread at a given galactocentric distance suggest that even a zone of 1~kpc$^2$ area has detectable metallicity spread. The measured metallicity of a zone corresponds to that of the brightest TRGBs in that zone, which in turn corresponds to high-end of the metallicity range at low extinction zones, and low metallicities at high extinction region. A comparison of the extrapolated RGB color with the bi-weighted median color for the simulated CMDs permits us to infer whether extinction spread or metallicity spread dictates the measured TRGB mag. We find that when extinction spread dominates, the extrapolated RGB color is redder than the bi-weighted median color, whereas it appears bluer when age-metallicity spread dominates. Figure~\ref{fig:col_diff} shows the distribution of the color difference between the extrapolated RGB color and the bi-weighted median color from the 90 regions. We observe that most regions exhibit a redder extrapolated RGB color, indicating that our regions have color spread arising predominantly from extinction spread rather than the age-metallicity spread.

\subsection{Implications for future TRGB studies using JWST CMDs}

This study has opened up a number of interesting problems that can be addressed using JWST images of nearby galaxies. First and foremost, we have demonstrated that large bubbles provide paths with little to no dust,  allowing for the determination of distances using the disk TRGB stars. The magnitude and colors of TRGBs allow a determination of spatially-resolved measurement of metallicities of the old stellar populations. Thus, the resolved stellar populations in the JWST filters provide a novel technique to determine stellar metallicities and their gradients at early times of disk formation.

\section{Conclusions}
\label{sec:conclusions}

By carrying out an analysis of the resolved stellar population in the JWST images of NGC\,628, we have demonstrated that the higher dependence of NIR TRGB magnitudes on metallicity and age as compared to that in the optical bands can be used to obtain metallicity gradients in the disks of galaxies without getting affected from the extinction. The near orthogonality of reddening and metallicity vectors in the F115W$-$F200W vs F200W diagram makes it possible to obtain the two quantities without degeneracy between these two quantities. The derived metallicities over 90~zones covering the entire FoV show a clear gradient. We obtain a median extinction of $A_V$= 0.12~mag in the disk caused by dust mixed with the diffuse gas. The TRGB magnitude obtained from the populations superposed on the recently discovered Phantom Void serves to set a distance scale to this galaxy as this bubble is devoid of interstellar material, thus providing a dust-free line of sight. We obtain a distance moduli of $29.80\pm0.05(stat)\pm0.06(sys)$ (PARSEC) and $29.83\pm0.05(stat)\pm0.06(sys)$ (BaSTI) from tilt-corrected TRGB measurement and $29.81\pm0.05(stat)\pm0.06(sys)$~mag (PARSEC) and 29.77$\pm0.05(stat)\pm0.06(sys)$~mag (BaSTI), using TRGB color-magnitude measurements, which is in agreement with that obtained recently using PNLF of DM=29.89$^{+0.06}_{-0.09}$~mag. The errors on the DM in our measurements are much smaller as compared to those in previous measurements using TRGBs on the HST images. These reduced errors in the measured TRGB magnitudes, along with the sensitivity of the NIR magnitudes and colors to metallicity variations, have allowed us to determine the spatially resolved metallicity of the oldest populations in the disk of NGC\,628. This study paves the way toward understanding the early growth of disks in galaxies using the resolved stellar populations in the JWST images.

\section{Acknowledgement}
We thank the referee for insightful comments that have improved the manuscript. ACk would like to thank the Secretaría de Ciencia, Humanidades, Tecnología e Innovación (SECIHTI) for his PhD grant. AB acknowledges  PRIN2022 (2022NEXMP8) "Radiative Opacities For Astrophysical Applications". We extend our gratitude to Dr. Ivanio Puerari and Dr. Manual Zamora for providing the OLINKI and Mixli clusters, respectively, at INAOE, which supported the computational requirements of this project.

This research has also utilized computational resources provided by El Laboratorio Nacional de Supercómputo del Sureste de México (LNS), under project no. \textsf{202404072C}.

This work is based on observations made with the NASA/ESA/CSA James Webb Space Telescope. The data were obtained from the Mikulski Archive for Space Telescopes at the Space Telescope Science Institute, operated by the Association of Universities for Research in Astronomy, Inc., under NASA contract NAS 5-03127 for JWST. These observations are associated with program \#1783 (JWST-FEAST; PI: Adamo Angela). The specific observations analyzed in this work can be accessed via doi:\dataset[10.17909/1zxh-fx32]{http://dx.doi.org/10.17909/1zxh-fx32}.

This research has utilized the NASA/IPAC Extragalactic Database (NED), operated by the Jet Propulsion Laboratory, California Institute of Technology, under contract with the National Aeronautics and Space Administration.

Additionally, this work incorporates observations taken as part of the PHANGS-MUSE large program \citep{phangs_muse}, as well as data products created from observations collected at the European Organisation for Astronomical Research in the Southern Hemisphere under ESO programs 1100.B-0651, 095.C-0473, and 094.C-0623 (PHANGS-MUSE; PI: Schinnerer); 094.B-0321 (MAGNUM; PI: Marconi); 099.B-0242, 0100.B-0116, 098.B-0551 (MAD; PI: Carollo); and 097.B-0640 (TIMER; PI: Gadotti). This research has also benefited from the ESO Science Archive Facility services.

\bibliography{references}{}
\bibliographystyle{aasjournal}

\end{document}